\documentclass[runningheads]{llncs}

\usepackage{color}
\usepackage[table]{xcolor}

\usepackage{style}

\usepackage{mathrsfs}
\usepackage{amsmath}
\usepackage{amssymb}

\usepackage{subcaption}
\usepackage[vlined]{algorithm2e}
\SetAlCapFnt{\normalfont}

\usepackage{longtable, array}

\usepackage{tikz}
\usetikzlibrary{shapes, chains, scopes, positioning,automata,decorations.pathreplacing,arrows.meta}

\usepackage{graphicx}
\usepackage{hyperref}
\usepackage{url}
\newcommand{\doi}[1]{\textsc{doi}: \href{http://dx.doi.org/#1}{\nolinkurl{#1}}}

\usepackage{arydshln}

\usepackage{colortbl}
\definecolor{Gray}{gray}{0.90}

\usepackage[firstpage]{draftwatermark}
\SetWatermarkText{\hspace*{5in}\raisebox{5.5in}{\includegraphics[scale=0.1]{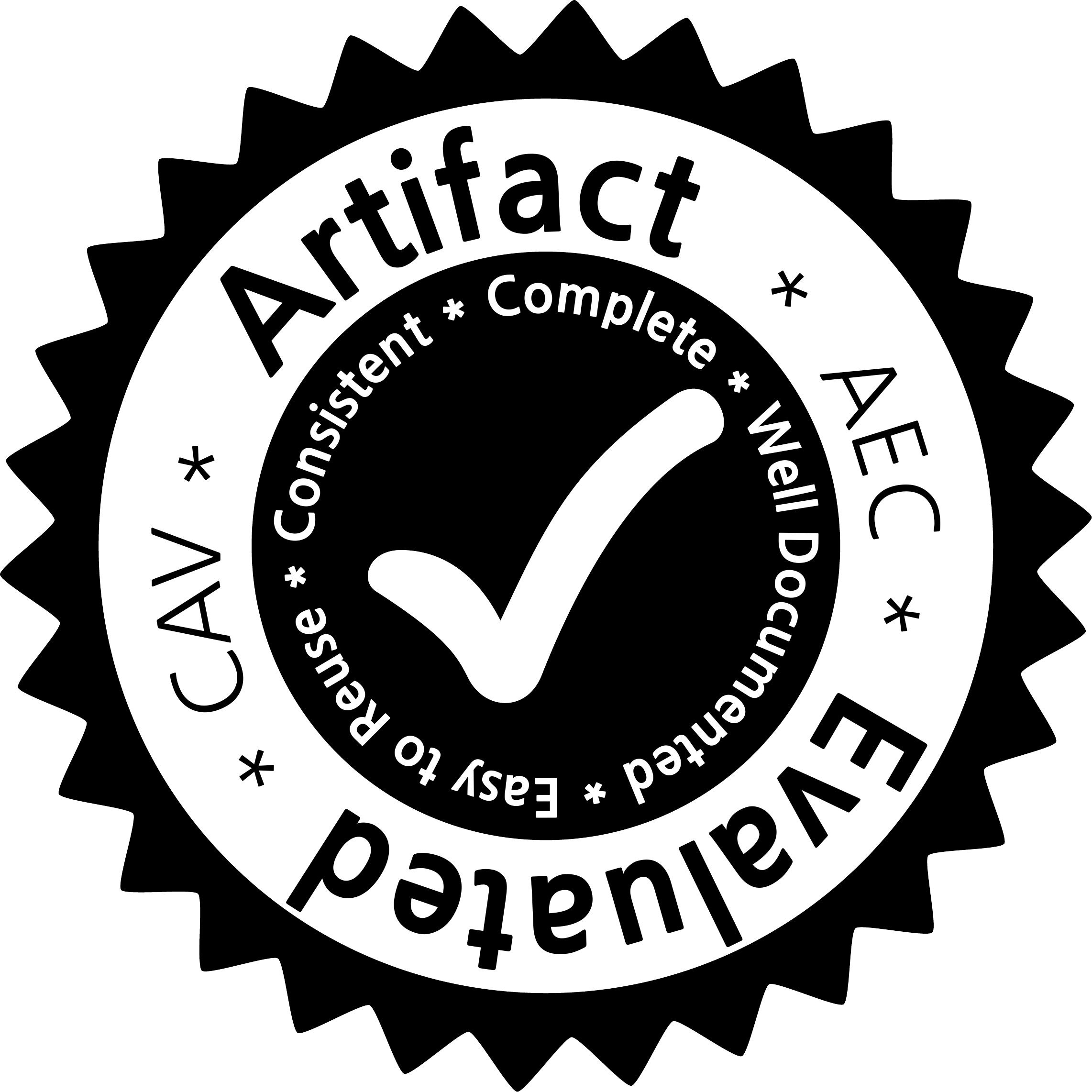}}}
\SetWatermarkAngle{0}

\usepackage[stable]{footmisc}

\begin{document}
\title{{\sc AdamMC}: A Model Checker for\\Petri Nets with Transits against Flow-LTL\\(Full Version)%
\thanks{This work was supported by the German Research Foundation (DFG) Grant Petri Games (392735815) and the Collaborative Research Center “Foundations of Perspicuous Software Systems” (TRR 248, 389792660), and by the European Research Council (ERC) Grant OSARES (683300).}%
\footnote[3]{This is the full version of \cite{DBLP:conf/cav/FinkbeinerGHO20}.}
}
\titlerunning{\adamMC: A Model Checker for Petri Nets with Transits against Flow-LTL}
%
\author{Bernd Finkbeiner\inst{1}
\and
Manuel Gieseking\inst{2}
\and\\
Jesko Hecking-Harbusch\inst{1}
\and
Ernst-R\"udiger Olderog\inst{2}
}

\authorrunning{B.\ Finkbeiner et al.}
%
\institute{Saarland University, Saarbr\"ucken, Germany\\
\email{\{finkbeiner,hecking-harbusch\}@react.uni-saarland.de} 
\and
University of Oldenburg, Oldenburg, Germany\\
\email{\{gieseking,olderog\}@informatik.uni-oldenburg.de}
}

\maketitle

\begin{abstract}
	The correctness of networks is often described in terms of the individual data flow of components instead of their global behavior. 
	In software-defined networks, it is far more convenient to specify the correct behavior of packets than the global behavior of the entire network. 
	Petri nets with transits extend Petri nets and Flow-LTL extends LTL such that the data flows of tokens can be tracked. 
	We present the tool \adamMC{} as the first model checker for Petri nets with transits against Flow-LTL. 
	We describe how \adamMC{} can automatically encode concurrent updates of software-defined networks as Petri nets with transits and how common network specifications can be expressed in Flow-LTL. 
	Underlying \adamMC{} is a reduction to a circuit model checking problem. 
	We introduce a new reduction method that results in tremendous performance improvements compared to a previous prototype. 
	Thereby, \adamMC{} can handle software-defined networks with up to 82 switches. 
\end{abstract}

\section{Introduction}
\label{sec:introduction}

In networks, it is difficult to specify correctness in terms of the global behavior of the entire system. 
Instead, the individual \emph{flow} of components is far more convenient to specify correct behavior. 
For example, loop and drop freedom can be easily specified for the flow of each packet.  
Petri nets and LTL lack this local view.
Petri nets with transits and Flow-LTL have been introduced to overcome this restriction~\cite{DBLP:conf/atva/FinkbeinerGHO19}. 
A transit relation is introduced to follow the \emph{flow} induced by tokens. 
\emph{Flow-LTL} is a temporal logic to specify both the \emph{local} flow of data and the \emph{global} behavior of markings. 
The global behavior as in Petri nets and LTL is still important for maximality and fairness assumptions. 
In this paper, we present the tool \adamMC{}\footnote{\adamMC{} is available online at \url{https://uol.de/en/csd/adammc}~\cite{tool}.} as the first model checker for Petri nets with transits against Flow-LTL and its application to software-defined networking. 

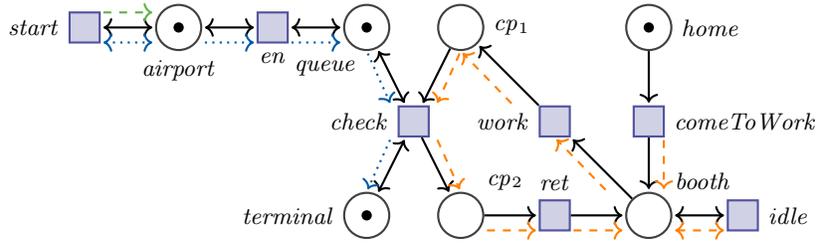
\begin{figure}[t]
 \centering
	\begin{tikzpicture}[node distance=1.25cm, on grid]%
		\node [envplace, tokens=1]						(s0)  [label=below:\textit{airport}] {};
		\node [transition, left=of s0] 					(t0)  [label=left:\textit{start}] {};
		\node [envplace, right=2.5cm of s0, tokens=1]	(s1)  [label=267:\textit{queue}] {};
		\node [transition, right=of s0] 					(t1)  [label=below:\textit{en}] {};
		\node [envplace, right=of s1]					(s2)  [label=right:\textit{cp}$_1$] {};
		\node [envplace, below=2.5cm of s1, tokens=1]	(s3)  [label=left:\textit{terminal}] {};
		\node [envplace, right=of s3]					(s4)  [label=45:\textit{cp}$_2$] {};
		\node [transition, below=of s1, xshift=0.625cm] 	(t6)  [label=left:\textit{check}] {}
			edge [pre] (s2)
			edge [post] (s4)
			;
		\node [envplace, right=2.5cm of s4]				(s5)  [label=45:\textit{booth}] {};
		\node [transition, right=of s4] 					(t3)  [label=above:\textit{ret}] {}
			edge [pre] (s4)
			edge [post] (s5);
		\node [transition, above=of t3] 					(t2)  [label=left:\textit{work}] {}
			edge [pre] (s5)
			edge [post] (s2);
		\node [envplace, right=2.5cm of s2, tokens=1]	(s6)  [label=right:\textit{home}] {};
		\node [transition, below=of s6] 					(t5)  [label=right:\textit{comeToWork}] {}
			edge [pre] (s6)
			edge [post] (s5);
		\node [transition, right=of s5] 					(t4)  [label=right:\textit{idle}] {};

		\draw[thick,shorten >=1pt,<->,shorten <=1pt] (s0) -- (t0);
		\draw[thick,<->, dotted, cdc_Blue, shorten >=1pt, shorten <=1pt] ([yshift=-0.2cm]t0.east) -- ([yshift=-0.2cm]s0.west);
		\draw[thick,->,dashed, cdc_Green, shorten >=1pt, shorten <=1pt] ([yshift=0.2cm]t0.east) -- ([yshift=0.2cm]s0.west);

		\draw[thick,shorten >=1pt,<->,shorten <=1pt] (s0) -- (t1);
		\draw[thick,shorten >=1pt,<->,shorten <=1pt] (t1) -- (s1);
		\draw[thick,->,dotted, cdc_Blue, shorten >=1pt, shorten <=1pt] ([yshift=-0.2cm]s0.east) -- ([yshift=-0.2cm]t1.west);
		\draw[thick,->,dotted, cdc_Blue, shorten >=1pt, shorten <=1pt] ([yshift=-0.2cm]t1.east) -- ([yshift=-0.2cm]s1.west);

		\draw[thick,shorten >=1pt,<->,shorten <=1pt] (s5) -- (t4);
		\draw[thick,<->,dashed, orange, shorten >=1pt, shorten <=1pt] ([yshift=-0.2cm]s5.east) -- ([yshift=-0.2cm]t4.west);

		\draw[thick,shorten >=1pt,<->,shorten <=1pt] (s1) -- (t6);
		\draw[thick,shorten >=1pt,<->,shorten <=1pt] (s3) -- (t6);

		\draw[thick,->,dashed, orange, shorten >=1pt, shorten <=1pt] (s2.south) -- ([xshift=0.3cm]t6.north);
		\draw[thick,->,dashed, orange, shorten >=1pt, shorten <=1pt] ([xshift=0.3cm]t6.south) -- (s4.north);
		\draw[thick,->,dotted, cdc_Blue, shorten >=1pt, shorten <=1pt] (s1.south) -- ([xshift=-0.3cm]t6.north);
		\draw[thick,->,dotted, cdc_Blue, shorten >=1pt, shorten <=1pt] ([xshift=-0.3cm]t6.south) -- (s3.north);

		\draw[thick,->,dashed, orange, shorten >=1pt, shorten <=1pt] ([xshift=-0.55cm]s5.north) -- (t2.south);
		\draw[thick,->,dashed, orange, shorten >=1pt, shorten <=1pt] ([xshift=-0.55cm]t2.north) -- (s2.south);

		\draw[thick,->,dashed, orange, shorten >=1pt, shorten <=1pt] ([yshift=-0.2cm]s4.east) -- ([yshift=-0.2cm]t3.west);
		\draw[thick,->,dashed, orange, shorten >=1pt, shorten <=1pt] ([yshift=-0.2cm]t3.east) -- ([yshift=-0.2cm]s5.west);

		\draw[thick,->,dashed, orange, shorten >=1pt, shorten <=1pt] ([xshift=0.2cm]t5.south) -- ([xshift=0.2cm]s5.north);

	\end{tikzpicture}
	\caption{
	Access control at an airport modeled as Petri net with transits.
	Colored arrows display the transit relation and define flow chains to model the passengers. 
	}
	\label{fig:motivatingexample}
\end{figure}

In \refFig{motivatingexample}, we present an example of a Petri net with transits that models the security check at an airport where passengers are checked by a security guard. 
The number of passengers entering the airport is unknown in advance.
Rather than introducing the complexity of an infinite number of tokens,
we use a fixed number of tokens to model possibly infinitely many \emph{flow chains}.
This is done by the transit relation which is depicted with colored arrows. 

The left-hand side of \refFig{motivatingexample} models passengers who want to reach the terminal.
There are three tokens in the places \textit{airport}, \textit{queue}, and \textit{terminal}.
Thus, transitions \textit{start} and \textit{en} are always enabled. 
Each firing of \textit{start} creates a new flow chain as depicted by the green arrow.
This models a new person arriving at the \textit{airport}.
Meanwhile, the double-headed blue arrow maintains all flow chains that are still in place \textit{airport}.
Passengers have to \textit{en}ter the \textit{queue} and wait until the security \textit{check} is performed.
Therefore, transition \textit{en} continues every flow chain in \textit{airport} to \textit{queue}.
Checking the passengers is carried out by transition \textit{check} which becomes enabled if the security guard \textit{work}s.
Thus, passengers residing in \textit{queue} have to wait until the guard \textit{check}s them.
Afterwards, they reach the \textit{terminal}. 
The security guard is modeled on the right-hand side of \refFig{motivatingexample}.
By firing \textit{comeToWork} and thus moving the token in place \textit{home}, her flow chain starts and she can repeatedly either \textit{idle} or \textit{work}, \textit{check} passengers, and \textit{ret}urn. 
Her transit relation is depicted in orange and models exactly one flow chain.

In \refFig{motivatingexample}, we define the checkpoints \textit{cp}$_1$ and \textit{cp}$_2$ and the \textit{booth} as a security zone and require that passengers never enter the security zone and eventually reach the \textit{terminal}. 
The flow formula
$\varphi = \A (airport \rightarrow (\LTLglobally \neg (\mathit{cp}_1 \lor \mathit{cp}_2 \lor \mathit{booth}) \land \LTLeventually terminal)) $ specifies this.
\adamMC{} verifies the example from \refFig{motivatingexample} against the formula $\LTLsquare \LTLeventually \mathit{check} \rightarrow \varphi$ specifying that if passengers are checked regularly then they cannot access the security zone and eventually reach the terminal. 

In this paper, we present \adamMC{} as a full-fledged tool.  
First, \adamMC{} can handle Petri nets with transits and Flow-LTL formulas in general. 
Second, \adamMC{} has an input interface for a concurrent update and a software-defined network and encodes both of them as a Petri nets with transits.
Common assumptions on fairness and requirements for network correctness are also provided as Flow-LTL formulas. 
This allows users of the tool to model check the correctness of concurrent updates and to prevent packet loss, routing loops, and network congestion.
Third, \adamMC{} provides algorithms to check safe Petri nets against LTL
with \emph{both} places and transitions as atomic propositions
which makes it especially easy to specify fairness and maximality assumptions.

The tool reduces the model checking problem for safe Petri nets with transits against Flow-LTL to the model checking problem for safe Petri nets against~LTL. 
We develop the new \emph{parallel approach} to check global and local behavior in parallel instead of sequentially. 
This approach yields a tremendous speed-up for a few local requirements and realistic fairness assumptions in comparison to the sequential approach of a previous prototype~\cite{DBLP:conf/atva/FinkbeinerGHO19}. 
In general, the parallel approach has worst-case complexity inferior to the sequential approach even though the complexities of both approaches are the same when using only one flow formula. 

As last step, \adamMC{} reduces the model checking problem of safe Petri nets against LTL to a circuit model checking problem.
This is solved by ABC~\cite{abcAsTheyWantIt,DBLP:conf/cav/BraytonM10} with effective verification techniques like IC3 and bounded model checking. 
\adamMC{} verifies concurrent updates of software-defined networks with up to 38 switches (31 more than the prototype) and falsifies concurrent updates of software-defined networks with up to 82 switches (44 more than the prototype).

The paper is structured as follows: In Sec.~\ref{sec:background}, we recall Petri nets with transits and Flow-LTL. 
In Sec.~\ref{sec:appAreas}, we outline the three application areas of \adamMC: checking safe Petri nets with transits against Flow-LTL, checking concurrent updates of software-defined networks against common assumptions and specifications, and checking safe Petri nets against LTL.
In Sec.~\ref{sec:sdn}, we algorithmically encode concurrent updates of software-defined networks in Petri nets with transits.
In Sec.~\ref{sec:algAndOpt}, we introduce the parallel approach for the underlying circuit model checking problem. 
In Sec.~\ref{sec:eval}, we present our experimental evaluation. 

\section{Petri Nets With Transits and Flow-LTL}
\label{sec:background}
A safe \emph{Petri net with transits} $\petriNetFl$~\cite{DBLP:conf/atva/FinkbeinerGHO19} contains the set of \emph{places}~$\pl$, the set of \emph{transitions}~$\tr$,
the \emph{flow relation}~$\fl \subseteq (\pl \times \tr) \cup (\tr \times \pl)$, and the \textit{initial marking}~$\init\subseteq\pl$ as in safe Petri nets~\cite{DBLP:books/sp/Reisig85a}. 
In a \emph{safe} Petri net, reachable markings contain at most one token per place. 
The \emph{transit relation}~$\tokenflow$ is for every transition $t \in \transitions$ of type $\tokenflow(t) \subseteq (\pre{\pNet}{t}\cup\{\startfl\}) \times \post{\pNet}{t}$. 
With~$p\ \tokenflow(t)\ q$, we define that firing transition~$t$ \emph{transits} the flow in place~$p$ to place~$q$.
The symbol~$\startfl$ denotes a \emph{start} and $\startfl\ \tokenflow(t)\ q$ defines that firing transition~$t$ \emph{starts} a new flow for the token in place~$q$. 
Note that the transit relation can split, merge, and end flows. 
A sequence of flows leads to a \emph{flow chain}  which is a sequence of the current place and the fired outgoing transition. 
Thus, Petri nets with transits can describe both the global progress of tokens and the local flow of data.

Flow-LTL~\cite{DBLP:conf/atva/FinkbeinerGHO19} extends Linear-time Temporal Logic (LTL) and uses places and transitions as atomic propositions. 
It introduces $\A$ as a new operator which uses LTL to specify the flow of data for \emph{all} flow chains. 
For \refFig{motivatingexample}, the formula $\A(\mathit{booth} \rightarrow \LTLeventually \mathit{check})$ specifies that the guard performs at least one check. 
We call formulas starting with $\A$ \emph{flow formulas}. 
Formulas around flow formulas specify the global progress of tokens in the form of markings and fired transitions to formalize maximality and fairness assumptions. 
These formulas are called \emph{run formulas}. 
Often, Flow-LTL formulas have the form $\mathit{run~formula} \rightarrow \mathit{flow~formula}$.

\section{Application Areas}
\label{sec:appAreas}
\adamMC{} consists of modules for three application areas:
checking safe Petri nets with transits against Flow-LTL,
checking concurrent updates of software-defined networks against common assumptions and specifications, 
and checking safe Petri nets against LTL.
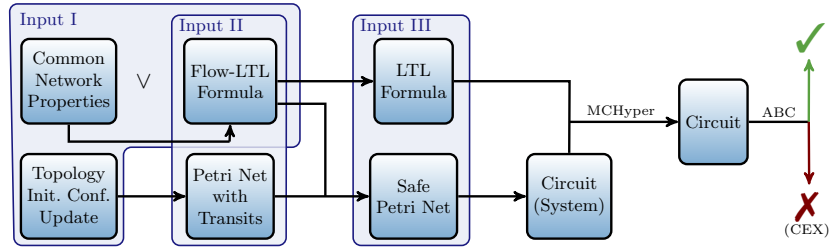
\begin{figure}[t]
\centering
 \scalebox{.75}{
 \begin{tikzpicture}[		
		->,
		very thick,
		>=stealth',
		node distance=5mm and 12mm,
		lbl/.style={
		   align=center
		},
		elem/.style={
			rectangle,
			rounded corners,
			draw=black, 
			very thick,
		    text centered,
			minimum height=10mm,
		}
 ]
	\tikzstyle{myarrows}=[line width=1mm,-triangle 45,postaction={line width=2mm, shorten >=2mm, -}]
	\node[elem, bottom color=cdc_Blue!50, top color=white,minimum height=15mm, minimum width=15mm, align=center, yshift=0mm] (props) {Common\\Network\\Properties};
	\node[elem, below=of props,bottom color=cdc_Blue!50, top color=white,minimum height=15mm, align=center, yshift=0mm] (sdn) {Topology\\Init. Conf.\\Update};
	\node[elem, right=of sdn, bottom color=cdc_Blue!50, top color=white,minimum height=15mm, align=center, yshift=0mm] (pnwt) {Petri Net\\with\\Transits};
	\node[elem,above=of pnwt, yshift=0mm, bottom color=cdc_Blue!50, minimum height=15mm, top color=white, align=center] (formula) {Flow-LTL\\Formula};
	\node[xshift=-1mm] at ($(props)!.5!(formula)$) {\large\(\mathbf{\vee}\)};
	\node[elem,right=of formula, bottom color=cdc_Blue!50, top color=white,minimum height=15mm, align=center, xshift=5mm] (f2) {LTL\\Formula};
	\node[elem,below=of f2, bottom color=cdc_Blue!50, align=center, top color=white,minimum height=15mm, yshift=0mm]  (pnmc) {Safe\\Petri Net};
	\node[elem,right=of pnmc, bottom color=cdc_Blue!50, top color=white,minimum height=15mm, align=center] (sysC) {Circuit\\(System)};
	\node[elem,right=of sysC, bottom color=cdc_Blue!50, top color=white,minimum height=15mm, xshift=60, yshift=3mm] (circuit) at ($(f2)!.5!(sysC)$) {Circuit};
    \node[label=below:{}, right=of circuit,xshift=-5mm,yshift=15mm] (yes) {\huge\color{cdc_Green!90!black}\cmark};
    \node[label={[yshift=2mm]below:{\scriptsize(CEX)}}, right=of circuit,xshift=-5mm,yshift=-15mm,align=center] (cex) {{\huge\color{ Red!70!black}\xmark}};

	\path (sdn) edge[very thick]  (pnwt);
	\path (pnwt) edge[very thick]  (pnmc);
	
    \draw[-, very thick] ([yshift=-4mm]formula.east) -- ++(8.5mm,0) -- ++(0,-16.5mm);
    \draw[-, very thick] (props.south) -- ++(0mm,-3mm) -- ++(28.5mm,0) edge[very thick]  (formula.south);
    
    \path (pnmc) edge [very thick]  (sysC);
    \path (formula) edge[very thick] (f2);
    
	\draw[-, very thick] (f2.east) -- ++(20.5mm, 0) -- (sysC.north);
    \path ([xshift=20.5mm]f2.east|-circuit.west)  edge[very thick]  node[above,yshift=-.5mm,xshift=-0.5mm] {\scriptsize{}MCHyper} (circuit.west);
    \draw[-, very thick] (circuit.east) -- node[above, yshift=-.5mm] {\scriptsize{}ABC} ($(cex)!.5!(yes)$);
    \path[cdc_Green!90!black, very thick,->] ($(cex.north)!.5!(yes.south)$) edge[very thick] ([xshift=-0.3mm]yes.south);
	\path[Red!70!black] ($(cex.north)!.5!(yes.south)$) edge[very thick] ([xshift=0.4mm]cex.north);

\node[xshift=5mm, yshift=3mm] at (props.north west) {\color{DarkBlue} Input I};
\node[xshift=5mm, yshift=2mm] at (formula.north west) {\color{DarkBlue} Input II};
\node[xshift=4mm, yshift=2mm] at (f2.north west) {\color{DarkBlue} Input III};

\begin{pgfonlayer}{background}
\draw [-, rectangle,rounded corners,DarkBlue,fill=cdc_BlueL!15]
 ([xshift=-1.9mm,yshift=6mm]props.north west) 
-- ([xshift=4mm,yshift=6mm]formula.north east)
-- ([xshift=4mm,yshift=-4mm]formula.south east)
-- ([xshift=1mm,yshift=1mm]sdn.north east)
-- ([xshift=1mm,yshift=-1mm]sdn.south east)
-- ([xshift=-1mm,yshift=-1mm]sdn.south west)
-- cycle;
\draw [-, rectangle,rounded corners,DarkBlue,fill=cdc_BlueL!15, fill opacity=0.8]
 ([xshift=-2mm,yshift=4mm]formula.north west) 
-- ([xshift=1.6mm,yshift=4mm]formula.north east)
-- ([xshift=2mm,yshift=-1mm]pnwt.south east)
-- ([xshift=-2.4mm,yshift=-1mm]pnwt.south west)
-- cycle;
\draw [-, rectangle,rounded corners,DarkBlue,fill=cdc_BlueL!15]
 ([xshift=-3.6mm,yshift=4mm]f2.north west) 
-- ([xshift=2.78mm,yshift=4mm]f2.north east)
-- ([xshift=2mm,yshift=-1mm]pnmc.south east)
-- ([xshift=-2.8mm,yshift=-1mm]pnmc.south west)
-- cycle;
\end{pgfonlayer}
\end{tikzpicture}
 }
	\caption{
		 Overview of the workflow of \(\adamMC\):
		 The application areas of the tool are given by three different input domains:
		 software-defined network / Flow-LTL (Input I), Petri nets with transits / Flow-LTL (Input II), and Petri nets / LTL (Input III).
		 \(\adamMC\) performs all unlabeled steps.
		 MCHyper creates the final circuit which ABC
		 checks to answer the initial model checking problem.
	}
	\label{fig:workflow}
\end{figure}
The general architecture and workflow of the model checking procedure is given in \refFig{workflow}.
\adamMC{} is based on the tool~\textsc{Adam}~\cite{DBLP:conf/cav/FinkbeinerGO15}.
\\\noindent\textbf{Petri Nets with Transits~~}
Petri nets with transits follow the progress of tokens and the flow of data. Flow-LTL allows to specify requirements on both. 
For Petri nets with transits and Flow-LTL (Input II),
\adamMC{} extends a parser for Petri nets provided by APT \cite{apt},
provides a parser for Flow-LTL,
and implements two reduction methods to create a safe Petri net and an LTL formula.
The sequential approach is outlined in \cite{DBLP:conf/atva/FinkbeinerGHO19} and the parallel approach 
in Sec.~\ref{sec:algAndOpt}.
\\\noindent\textbf{Software-Defined Networks~~}
Concurrent updates of software-defined networks are the second application area of \adamMC. 
The tool automatically encodes an initially configured network topology and a concurrent update as a Petri net with transits. 
The concurrent update renews the forwarding table. 
We provide parsers for
the \emph{network topology}, the \emph{initial configuration}, the \emph{concurrent update},
and Flow-LTL (Input~I).
In Sec.~\ref{sec:sdn}, we present the creation of a Petri net with transits from the input and Flow-LTL formulas for \emph{common network properties} like
\emph{connectivity}, \emph{loop freedom}, \emph{drop freedom}, and \emph{packet coherence}.
\\\noindent\textbf{Petri Nets~~}
\adamMC{} supports the model checking of safe Petri nets against LTL with both places \emph{and} transitions as 
atomic propositions.
It provides dedicated algorithms to check \emph{interleaving-maximal} 
runs of the system. 
A run is interleaving-maximal if a transition is fired whenever a transition is enabled.
Furthermore, \adamMC{} allows a concurrent view on runs and can check 
\emph{concurrency-maximal} runs which demand that each subprocess of the system has to progress maximally
rather than only the entire system.
State-of-the-art tools like LoLA~\cite{DBLP:conf/apn/Wolf18a} and ITS-Tools~\cite{DBLP:conf/tacas/Thierry-Mieg15}
are restricted to interleaving-maximal runs and places as atomic propositions.
For Petri net model checking (Input III), 
we allow Petri nets in APT and PNML format as input and provide a parser for LTL formulas.

The construction of the circuit in Aiger format \cite{Biere-FMV-TR-11-2} is defined in \cite{DBLP:journals/corr/abs-1907-11061}.
MCHyper~\cite{DBLP:conf/cav/FinkbeinerRS15} is used to create a circuit from a given circuit and an LTL formula.
This circuit is given to ABC \cite{abcAsTheyWantIt,DBLP:conf/cav/BraytonM10} which provides a toolbox of modern hardware
verification algorithms like IC3 and bounded model checking to decide the initial model checking question.
As output for all three modules, \adamMC{} transforms a possible counterexample (CEX) from ABC into
a counterexample to the Petri net (with transits) 
and visualizes the net with Graphviz and the dot language~\cite{DBLP:books/sp/04/EllsonGKNW04}. 
When no counterexample exists, \adamMC{} verified the input successfully.

\section{Verifying Updates of Software Defined Networks}
\label{sec:sdn}
We show how \adamMC{} can check concurrent updates of realistic examples from software-defined networking (SDN) against typical specifications~\cite{DBLP:journals/jsac/KnightNFBR11}. 
SDN~\cite{DBLP:journals/ccr/McKeownABPPRST08,DBLP:journals/cacm/CasadoFG14} separates the \emph{data plane} for forwarding packets and the \emph{control plane} for the routing configuration.
A central controller initiates updates which can cause problems like routing loops or packet loss. 
\adamMC{} provides an input interface to automatically encode software-defined networks and concurrent updates of their configuration as Petri nets with transits. 
The tool checks requirements like loop and drop freedom to find erroneous updates before they are deployed. 

\subsection{Network Topology, Configurations, and Updates}

A \emph{network topology} $T$ is an undirected graph $T=(\mathit{Sw}, \mathit{Con})$ with \emph{switches} as vertices and \emph{connections} between switches as edges.
Packets enter the network at \emph{ingress} switches and they leave at \emph{egress} switches.
\emph{Forwarding} rules are of the form $\texttt{x.fwd(y)}$ with $\texttt{x}, \texttt{y} \in \mathit{Sw}$.
A concurrent \emph{update} has the following syntax:
\[
\begin{array}{lll}
	\text{switch update} &::= \texttt{upd(x.fwd(y/z))\,} | \texttt{\,upd(x.fwd(y/-))\,} | \texttt{\,upd(x.fwd(-/z))}\\
	\text{sequential update} &::= \texttt{(}\text{update } \texttt{>>} \text{ update } \texttt{>>} \text{ ... } \texttt{>>} \text{ update}\texttt{)} \\
	\text{parallel update} &::= \texttt{(} \text{update } \texttt{||} \text{ update } \texttt{||} \text{ ... } \texttt{||} \text{ update} \texttt{)} \\
	\text{update} &::= \text{switch update\,} | \text{\,sequential update\,} | \text{\,parallel update}\\
\end{array}
\]
where a switch update can renew the forwarding rule of switch~\texttt{x} from switch~\texttt{z} to switch~\texttt{y}, introduce a new forwarding rule from switch~\texttt{x} to switch~\texttt{y}, or remove an existing forwarding rule from switch~\texttt{x} to switch~\texttt{z}. 

\subsection{Data Plane and Control Plane as Petri Net with Transits}
\label{subsec:motivation-data-plane}

For a network topology $T = (\mathit{Sw}, \mathit{Con})$, a set of \textit{ingress} switches, a set of \textit{egress} switches, an initial \textit{forwarding} table,  
and a concurrent $\mathit{update}$, we show how data and control plane are encoded as Petri net with transits.
Switches are modeled by tokens remaining in corresponding places~\texttt{s} whereas the flow of packets is modeled by the transit relation~$\tfl$. 
Specific transitions~$i_\texttt{s}$ model ingress switches where new data flows begin.
Tokens in places of the form \texttt{x.fwd(y)} configure the forwarding. 
Data flows are extended by firing transitions~$\texttt{(x,y)}$ corresponding to configured forwarding without moving any tokens.
Thus, we model any order of newly generated packets and their forwarding.
Assuming that each existing direction of a connection between two switches is explicitly given in $\mathit{Con}$, 
we obtain Algorithm~\ref{alg:data-plane} which calls Algorithm~\ref{alg:control-plane} to obtain the control plane. 

\hspace*{-10mm}
\begin{minipage}[t]{0.5\textwidth}
	\begin{algorithm}[H]
	\SetVlineSkip{0.25mm}
	\footnotesize
 	\SetKwInOut{Input}{input}\SetKwInOut{Output}{output}
 	\Input{$T=(\mathit{Sw},\mathit{Con})$, \textit{ingress}, \textit{forwarding}, \textit{update}}
 	\Output{Petri net with transits $\petriNetFl$ for \textit{update} of topology $T$ with \textit{ingress} and \textit{forwarding}}
 	create empty\,$\petriNetFl$\;
 	\For{\emph{switch \texttt{s}} $\in \mathit{Sw}$} {
 		add place \texttt{s} to $\pl$\;
 	 	add place \texttt{s} to $\init$\;
 	 	\If{\emph{\texttt{s}} $\in$ ingress}{
 	 	 add transition $i_\texttt{s}$ to $\tr$\;
 	 	 add \texttt{s} to $\pre{}{i_\texttt{s}}$, $\post{}{i_\texttt{s}}$\;
 	 	 add creating data flow $\startfl~\tokenflow(i_\texttt{s})~\texttt{s}$ to $\tfl$\;
 	 	 add maintaining data flow $\texttt{s}~\tokenflow(i_\texttt{s})~\texttt{s}$ to $\tfl$\;
 	 	}
 	}
 	\For{\emph{connection} $(\emph{\texttt{x}},\emph{\texttt{y}}) \in \mathit{Con}$} {
 		add place $\texttt{x.fwd(y)}$ to $\pl$\;
 		\If{$\emph{\texttt{x.fwd(y)}} \in$ forwarding}{
 	 	 	add place $\texttt{x.fwd(y)}$ to $\init$\;
 	 	}
 		add transition $(\texttt{x},\texttt{y})$ to $\tr$\;
 		add \texttt{x}, \texttt{y}, \texttt{x.fwd(y)} to $\pre{}{(\texttt{x},\texttt{y})}$, $\post{}{(\texttt{x},\texttt{y})}$\;
 		add connecting data flow $\texttt{x}~\tokenflow((\texttt{x},\texttt{y}))~\texttt{y}$ to $\tfl$\;
 		add maintaining data flow $\texttt{y}~\tokenflow((\texttt{x},\texttt{y}))~\texttt{y}$ to $\tfl$\;
 	}
 	$\pNet =$ call Algorithm~\ref{alg:control-plane} with $T$, \textit{update}, $\pNet$ as input\;
 	add place $\mathit{update}^s$ to $\init$\;
 	\vspace{0.035cm}
	\caption{Data plane}
	\label{alg:data-plane}
	\end{algorithm}
\end{minipage}
\hfill
\begin{minipage}[t]{0.6\textwidth}
\begin{algorithm}[H]
	\SetVlineSkip{0.25mm}
 	\footnotesize
 	\SetKwInOut{Input}{input}\SetKwInOut{Output}{output}
 	\Input{$T=(\mathit{Sw},\mathit{Con})$, \textit{update}, $\pNet$}
 	\Output{$\petriNetFl$}
 	\For{\emph{switch update} $u\in \mathit{SwU}$} {
 		// $u = \texttt{upd(x.fwd(y/z))}$\\
 		add places $u^s$, $u^f$ to $\pl$\;
 		add transition $u$ to $\tr$\;
 		add $u^s$ to $\pre{}{u}$, $u^f$ to $\post{}{u}$\;
 		\If{$\texttt{z} \neq \texttt{-}$}{
 	 	 	add $\texttt{x.fwd(z)}$ to $\pre{}{u}$\;
 	 	}
 	 	\If{$\texttt{y} \neq \texttt{-}$}{
 	 	 	add $\texttt{x.fwd(y)}$ to $\post{}{u}$\;
 	 	}
 	}
 	\For{\emph{sequential update} $s \in \mathit{SeU}$} {
 		// $s = [s_1,...,s_i,...,s_{|s|}]$ \\
 		add places $s^s$, $s^f$ to $\pl$\;
 		\For{$i \in \{0,...,|s|\}$} {
 			add transition $s^i$ to $\tr$\;
 			\uIf{$i == 0$}{
 	 	 		add $s^s$ to $\pre{}{s^i}$\;
 	 		}
 	 		\Else{
 	 			add $s^f_i$ to $\pre{}{s^i}$\;
 	 		}
 	 		\uIf{$i = |s|$}{
 	 	 		add $s^f$ to $\post{}{s^i}$\;
 	 		}
 	 		\Else{
 	 			add $s_{i+1}^s$ to $\post{}{s^i}$\;
 	 		}
 		}
 	}
 	\For{\emph{parallel update} $p \in \mathit{PaU}$} {
 		add places $p^s$, $p^f$ to $\pl$\;
 		add transitions $p^o$, $p^c$ to $\tr$\;
 		add $p^s$ to $\pre{}{p^o}$, $p^f$ to $\post{}{p^c}$\;
 		\For{\emph{sub-update} $u_i$ \emph{of} $p$} {
 			add $u^s_i$ to $\post{}{p^o}$, $u^f_i$ to $\pre{}{p^c}$\;
 		}
 	}
 	\caption{Control plane}
 	\label{alg:control-plane}
	\end{algorithm}
\end{minipage}

For the $\mathit{update}$, let $\mathit{SwU}$ be the set of switch updates in it, $\mathit{SeU}$ the set of sequential updates in it, and $\mathit{PaU}$ the set of parallel updates in it. 
Depending on $\mathit{update}$'s type, it is also added to the respective set.
The subnet for the \textit{update} has an empty transit relation but moves tokens from and to places of the form $\texttt{x.fwd(y)}$. 
Tokens in these places correspond to the forwarding table. 
The order of the switch updates is defined by the nesting of sequential and parallel updates.
The \textit{update} is realized by a specific token moving through unique places of the form $u^s, u^f, s^s, s^f, p^s, p^f$ for start and finish of each switch update $u\in\mathit{SwU}$, each sequential update $s\in\mathit{SeU}$, and each parallel update $p\in\mathit{PaU}$.
A parallel update temporarily increases the number of tokens and reduces it upon completion to one.
Algorithm~\ref{alg:control-plane} defines the update behavior between start and finish places and connects finish and start places depending on the subexpression structure.

\subsection{Assumptions and Requirements}

We use the run formula $\LTLeventually\LTLglobally \pre{}{t} \rightarrow \LTLglobally\LTLeventually t$ to assume weak fairness for every transition~$t$ in our encoding~$\pNet$. 
Transitions, which are always enabled after some point, are ensured to fire infinitely often. 
Thus, packets are eventually forwarded and the routing table is eventually updated.   
We use flow formulas to test specific requirements for all packets.
Connectivity ($\A (\LTLeventually \bigvee_{s\in\mathit{egress}} s) $) ensures that all packets reach an egress switch.
Packet coherence ($\A ( \ltlAlways\, ( \bigvee_{\texttt{s}\in\textit{initial}}\texttt{s}) \lor \ltlAlways\, (\bigvee_{\texttt{s}\in\textit{final}}\texttt{s}) )$) tests that packets are either routed according to the initial or final configuration. 
Drop freedom ($\A\ltlAlways\, (\bigwedge_{\texttt{e}\in\mathit{egress}}\neg \texttt{e} \rightarrow \bigvee_{f\in\mathit{Con}} f)$) forbids dropped packets whereas loop freedom ($\A \ltlAlways\, (\bigwedge_{\texttt{s}\in\mathit{Sw}\setminus\mathit{egress}} \texttt{s} \rightarrow (\texttt{s} \U \ltlAlways \neg \texttt{s}))$) forbids routing loops.
We combine run and flow formula into \emph{fairness} $\rightarrow$ \emph{requirement}.

\section{Algorithms and Optimizations}
\label{sec:algAndOpt}
Central to model checking a Petri net with transits \(\pNet\)~against a Flow-LTL formula \(\phiRun\)
is the reduction to a safe Petri net \(\pNetMC\) and an LTL formula \(\phiMC\).
The infinite state space of the Petri net with transits due to possibly infinitely many flow chains
is reduced to a finite state model.
The key idea is to guess and track a violating flow chain for each flow subformula \(\A\,\phiLTL_i\), for \(i\in\{1,\ldots,n\}\),
and to only once check the equivalent future of flow chains merging into a common place.

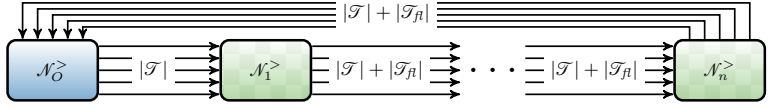
\begin{figure}[t]
\centering
 \scalebox{.8}{
 \begin{tikzpicture}[		
		->,
		thick,
		>=stealth',
		node distance=25mm and 20mm,
		lbl/.style={
		   align=center
		},
		elem/.style={
			rectangle,
			rounded corners,
			draw=black, 
			very thick,
		    text centered,
			minimum height=5mm,
		}
 ]
	\tikzstyle{myarrows}=[line width=1mm,-triangle 45,postaction={line width=2mm, shorten >=2mm, -}]
	\node[elem, bottom color=cdc_Blue!50, top color=white,minimum height=10mm,,minimum width=15mm, align=center, yshift=0mm] (o) {\(\pNetMCO\)};
	\node[elem, right=of o, fill opacity=0.95,pattern=checkerboard, pattern color=cdc_Green!50!black, bottom color=cdc_Green!50, top color=white,minimum height=10mm,minimum width=15mm, align=center, yshift=0mm] (sub1) {\(\pNetMCSub{1}\)};
	\node[elem, right=of sub1, fill opacity=0.95,pattern=checkerboard, pattern color=cdc_Green!50!black, bottom color=cdc_Green!50, top color=white,minimum height=10mm,minimum width=15mm, align=center, xshift=40mm] (subn) {\(\pNetMCSub{n}\)};

\node[rectangle, fill=white,minimum height=10mm,minimum width=6mm, align=center, xshift=0mm] at ($(sub1)!.5!(subn)$) (dots) {\huge\(\dots\)};

	\path (o) edge (sub1);
	\path ([yshift=2mm]o.east) edge ([yshift=2mm]sub1.west);
	\path ([yshift=4mm]o.east) edge ([yshift=4mm]sub1.west);
	\path ([yshift=-2mm]o.east) edge ([yshift=-2mm]sub1.west);
	\path ([yshift=-4mm]o.east) edge ([yshift=-4mm]sub1.west);

	\path (sub1) edge (dots);
	\path ([yshift=2mm]sub1.east) edge ([yshift=2mm]dots.west);
	\path ([yshift=4mm]sub1.east) edge ([yshift=4mm]dots.west);
	\path ([yshift=-2mm]sub1.east) edge ([yshift=-2mm]dots.west);
	\path ([yshift=-4mm]sub1.east) edge ([yshift=-4mm]dots.west);

	\path (dots) edge (subn);
	\path ([yshift=2mm]dots.east) edge ([yshift=2mm]subn.west);
	\path ([yshift=4mm]dots.east) edge ([yshift=4mm]subn.west);
	\path ([yshift=-2mm]dots.east) edge ([yshift=-2mm]subn.west);
	\path ([yshift=-4mm]dots.east) edge ([yshift=-4mm]subn.west);

	\node[rectangle, fill=white,minimum height=6mm,minimum width=6mm, align=center, xshift=-1mm] at ($(o)!.5!(sub1)$) (trO-1) {\(|\tr|\)};
	\node[rectangle, fill=white,minimum height=6mm,minimum width=6mm, align=center, xshift=0mm] at ($(sub1)!.5!(dots)$) (trO-2) {\(|\tr|+|\tr_\mathit{fl}|\)};
	\node[rectangle, fill=white,minimum height=6mm,minimum width=6mm, align=center, xshift=-2mm] at ($(dots)!.5!(subn)$) (trO-3) {\(|\tr|+|\tr_\mathit{fl}|\)};
	
	\draw[->] (subn.north) -- ++(0,4mm) -- ([yshift=4mm]o.north) -- (o.north);
	\draw[->] ([xshift=-2.5mm]subn.north) -- ++(0,3mm) -- ([yshift=3mm, xshift=2.5mm]o.north) -- ([xshift=2.5mm]o.north);
	\draw[->] ([xshift=-5mm]subn.north) -- ++(0,2mm) -- ([yshift=2mm, xshift=5mm]o.north) -- ([xshift=5mm]o.north);
	\draw[->] ([xshift=2.5mm]subn.north) -- ++(0,5mm) -- ([yshift=5mm, xshift=-2.5mm]o.north) -- ([xshift=-2.5mm]o.north);
	\draw[->] ([xshift=5mm]subn.north) -- ++(0,6mm) -- ([yshift=6mm, xshift=-5mm]o.north) -- ([xshift=-5mm]o.north);

	\node[rectangle, fill=white,minimum height=2mm,minimum width=2mm, align=center, yshift=9.5mm, xshift=0mm] at ($(o)!.5!(subn)$) (trO-3) {\small\(|\tr|+|\tr_\mathit{fl}|\)};

\end{tikzpicture}
 }
	\caption{Overview of the \emph{sequential approach}: Each firing of a transition of the original net
is split into first firing a transition in the subnet for the run formula
and subsequently firing a transition in each subnet tracking a flow formula.
The constructed LTL formula skips the additional steps with until operators.
\vspace{-0.2cm}}
	\label{fig:overviewMCSeq}
\end{figure}
\begin{figure}[t]
\centering
 \scalebox{.8}{
 \begin{tikzpicture}[		
		->,
		thick,
		>=stealth',
		node distance=25mm and 20mm,
		lbl/.style={
		   align=center
		},
		elem/.style={
			rectangle,
			rounded corners,
			draw=black, 
			very thick,
		    text centered,
			minimum height=10mm,
		}
 ]
	\tikzstyle{myarrows}=[line width=1mm,-triangle 45,postaction={line width=2mm, shorten >=2mm, -}]
	\node[elem, bottom color=cdc_Blue!50, top color=white,minimum height=10mm,,minimum width=15mm, align=center, yshift=0mm] (o) {\(\pNetMCO\)};
	\node[elem, right=of o, fill opacity=0.95,pattern=checkerboard, pattern color=cdc_Green!50!black, bottom color=cdc_Green!50, top color=white,minimum height=10mm,minimum width=15mm, align=center, yshift=0mm] (sub1) {\(\pNetMCSub{1}\)};
	\node[elem, right=of sub1, fill opacity=0.95,pattern=checkerboard, pattern color=cdc_Green!50!black, bottom color=cdc_Green!50, top color=white,minimum height=10mm,minimum width=15mm, align=center, xshift=40mm] (subn) {\(\pNetMCSub{n}\)};
\node[rectangle, fill=white,minimum height=10mm,minimum width=6mm, align=center, xshift=0mm] at ($(sub1)!.5!(subn)$) (dots) {\huge\(\mathbf{\dots}\)};

	\draw[-] (subn.south) -- ++(0,-4mm) -- ([yshift=-4mm]o.south) -- (o.south);
	\draw[-] ([xshift=-2.5mm]subn.south) -- ++(0,-3mm) -- ([yshift=-3mm, xshift=2.5mm]o.south) -- ([xshift=2.5mm]o.south);
	\draw[-] ([xshift=-5mm]subn.south) -- ++(0,-2mm) -- ([yshift=-2mm, xshift=5mm]o.south) -- ([xshift=5mm]o.south);
	\draw[-] ([xshift=2.5mm]subn.south) -- ++(0,-5mm) -- ([yshift=-5mm, xshift=-2.5mm]o.south) -- ([xshift=-2.5mm]o.south);
	\draw[-] ([xshift=5mm]subn.south) -- ++(0,-6mm) -- ([yshift=-6mm, xshift=-5mm]o.south) -- ([xshift=-5mm]o.south);

	\draw[-] (sub1.south) -- ++(0,-4mm);
	\draw[-] ([xshift=-2.5mm]sub1.south) -- ++(0,-3mm);
	\draw[-] ([xshift=-5mm]sub1.south) -- ++(0,-2mm) ;
	\draw[-] ([xshift=2.5mm]sub1.south) -- ++(0,-5mm) ;
	\draw[-] ([xshift=5mm]sub1.south) -- ++(0,-6mm) ;

	\node[rectangle, fill=white,minimum height=2mm,minimum width=2mm, align=center, yshift=-9.5mm, xshift=0mm] at ($(sub1)!.5!(subn)$) (trO-3) {\small\(|\tr|\cdot(|\tr_\mathit{fl}|+1)^n\)};

\end{tikzpicture}
 }
	\caption{Overview of the \emph{parallel approach}: The \(n\)~subnets are connected such that
for every transition \(t\in\tr\) there are \((|\tokenflow(t)|+1)^n\) transitions, i.e.,
there is one transition for every combination of which transit of \(t\) (or none) is tracked by which subnet.
We use until operators in the constructed LTL formula to only skip steps not involving the tracking of the guessed chain
in the flow formula.
\vspace{-0.2cm}}
	\label{fig:overviewMCPar}
\end{figure}
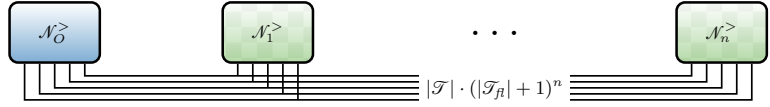

\adamMC{} provides two approaches for this reduction:
Fig.~\ref{fig:overviewMCSeq} and Fig.~\ref{fig:overviewMCPar} give
an overview of the \emph{sequential} approach and the \emph{parallel} approach, respectively.
Both algorithms create one subnet~\(\pNetMCSub{i}\) for each flow subformula~\(\A\,\phiLTL_i\) to track the corresponding flow chain
and have one subnet~\(\pNetMCO\) to check the run part of the formula.
The places of~\(\pNetMCO\) are copies of the places in~\(\pNet\) such that the current state of the system can be memorized.
The subnets~\(\pNetMCSub{i}\) also consist of the original places of~\(\pNet\)
but only use one token (initially residing on an additional place)
to track the current state of the considered flow chain.
The approaches differ in how these nets are connected to obtain~\(\pNetMC\).\\
\noindent\textbf{Sequential Approach~~}
The places in each subnet~\(\pNetMCSub{i}\) are connected 
with one transition for each transit (\(\tr_\mathit{fl}=\bigcup_{t\in\tr} \tokenflow(t)\)).
An additional token iterates sequentially through
the subnets to activate or deactivate the subnet.
This allows each subnet to track a flow chain
corresponding to firing a transition in~\(\pNetMCO\).
The formula \(\phiMC\) takes care of these additional steps
by means of the until operator:
In the run part of the formula, all steps corresponding to moves in a subnet \(\pNetMCSub{i}\) are skipped
and, for each subformula \(\A\,\phiLTL_i\), all steps are skipped until
the next transition of the corresponding subnet is fired which transits the tracked flow chain.
This technique results in a polynomial increase of the size of the Petri net and the formula:
\(\pNetMC\) has \(\oclass{|\pNet|\cdot n + |\pNet|}\) places 
and \(\oclass{|\pNet|^3\cdot n + |\pNet|}\) transitions and the size of \(\phiMC\)
is in \(\oclass{|\pNet|^3\cdot n \cdot |\phiRun| + |\phiRun|}\).
We refer to \cite{DBLP:journals/corr/abs-1907-11061} for formal details.\\
\noindent\textbf{Parallel Approach~~} The \(n\)~subnets are connected such that the current chain
of each subnet is tracked simultaneously while firing an original transition \(t\in\tr\).
Thus, there are \((|\tokenflow(t)|+1)^n\) transitions.
Each of these transitions stands for exactly one combination of
which subnet is tracking which (or no) transit.
Hence, firing one transition of the original net is directly tracked in one step for all subnets.
This significantly reduces the complexity of the run part of the constructed formula,
because no until operator is needed to skip sequential steps.
A disjunction over all transitions corresponding to an original transition suffices to 
ensure correctness of the construction. 
Transitions and next operators in the flow parts of the formula still have to be replaced
by means of the until operator to ensure that the next step of the tracked flow chain 
is checked at the corresponding step of the global timeline of \(\phiMC\).
In general, the parallel approach results in an exponential blow-up of the net and the formula:
\(\pNetMC\) has \(\oclass{|\pNet|\cdot n + |\pNet|}\) places 
and \(\oclass{|\pNet|^{3n}+|\pNet|}\) transitions and the size of \(\phiMC\)
is in \(\oclass{|\pNet|^{3n}\cdot |\phiRun| + |\phiRun|}\).
For the practical examples, however, the parallel approach allows for model checking Flow-LTL with few flow subformulas
with a tremendous speed-up in comparison to the sequential approach.
We refer to App.~\ref{app:parApp} for formal details. 
\\\noindent\textbf{Optimizations~~}
Various optimizations parameters
can be applied to the model checking routine described in Sec.~\ref{sec:appAreas}
to tweak the performance. Table~\ref{table:optimisations} gives an overview of the major parameters.
\bgroup
\def\arraystretch{1.1}%
\setlength\tabcolsep{1.2mm}
\begin{table}[t]
\caption{Overview of optimization parameters of \adamMC:
The three reduction steps depicted in the first column can each be executed by different algorithms.
The first step allows to combine the optimizations of the first and second row.}
  \label{table:optimisations}
  \centering
\begin{tabular}{|>{\columncolor{gray!50}}l|c|c|c|c|}\hline
1) Petri Net with Transits \(\leadsto\) Petri Net & \multicolumn{2}{c|}{sequential} & \multicolumn{2}{c|}{parallel} \\\cline{2-5}
								 & inhibitor & act.\ token &  inhibitor & act.\ token \\\hline
2) Petri Net  \(\leadsto\) Circuit & \multicolumn{2}{c|}{explicit} & \multicolumn{2}{c|}{logarithmic}  \\\hline
3) Circuit \(\leadsto\) Circuit & \multicolumn{4}{c|}{gate optimizations}\\\hline
\end{tabular}
\end{table}
\egroup
We found that the versions of the sequential and the parallel approach with
inhibitor arcs to track flow chains are generally faster than the versions without.
Furthermore, the reduction step from a Petri net into a circuit with logarithmically encoded
transitions had oftentimes better performance than the same step with explicitly encoded transitions. 
However, several possibilities to reduce the number of gates of the created circuit
worsened the performance of some benchmark families and improved the performance of others.
Consequently, all parameters are selectable by the user and
a script is provided to compare different settings.
An overview of the selectable optimization parameters can be found in the documentation of \adamMC{}~\cite{tool}.
Our main improvement claims
can be retraced by the case study in Sec.~\ref{sec:eval}.

\section{Evaluation}
\label{sec:eval}
{
\setlength{\tabcolsep}{1pt}
\begin{table}[t]
\caption{We compare the explicit and logarithmic encoding of the sequential approach with the parallel approach. The results are the average over five runs from an Intel i7-2700K CPU with 3.50~GHz, 32~GB RAM, and a timeout (TO) of 30~minutes. The runtimes are given in seconds.
}
  \label{table:benchmarks}
  \centering
\begin{tabular}{ccc|crc|crc|crc}
 &  & & \multicolumn{3}{c|}{expl.\ enc.} & \multicolumn{3}{c|}{log.\ enc.} & \multicolumn{3}{c}{parallel appr.}\\
T / F & Network & \(\#\mathit{Sw}\) & Alg. & Time & \(\models\) & Alg. & Time & \(\models\) & Alg. & Time & \(\models\)\\
 \cline{1-12}
T & Arpanet196912 &4& IC3 & 12.08 & \cmark & IC3 & 9.89 & \cmark & IC3 & \textbf{2.18} & \cmark\\
T & Napnet &6& IC3 & 146.49 & \cmark & IC3 & 96.06 & \cmark & IC3 & \textbf{4.75} & \cmark\\
&\(\cdots\)& & \multicolumn{3}{c|}{\(\cdots\)} & \multicolumn{3}{c|}{\(\cdots\)} & \multicolumn{3}{c}{\(\cdots\)}\\
T & Heanet &7& IC3 & 806.81 & \cmark & IC3 & 84.62 & \cmark & IC3 & \textbf{30.30} & \cmark\\
T & HiberniaIreland &7& - & TO & ? & - & TO & ? & IC3 & \textbf{26.58} & \cmark\\
T & Arpanet19706 &9& - & TO & ?& IC3 & 362.21 & \cmark  & IC3 & \textbf{11.33} & \cmark\\
T & Nordu2005 &9& - & TO & ? & - & TO & ? & IC3 & \textbf{12.67} & \cmark\\
&\(\cdots\)& & \multicolumn{3}{c|}{\(\cdots\)} & \multicolumn{3}{c|}{\(\cdots\)} & \multicolumn{3}{c}{\(\cdots\)}\\
T & Fatman &17& - & TO & ? & IC3 & 1543.34 & \cmark & IC3 & \textbf{162.17} & \cmark\\
&\(\cdots\)& & \multicolumn{3}{c|}{\(\cdots\)} & \multicolumn{3}{c|}{\(\cdots\)} & \multicolumn{3}{c}{\(\cdots\)}\\
T & Myren &37& - & TO & ? & - & TO & ? & IC3 & \textbf{1309.23} & \cmark\\
T & KentmanJan2011 &38& - & TO & ? & - & TO & ? & IC3 & \textbf{1261.32} & \cmark\\
\cline{1-12}
F & Arpanet196912  &4& BMC3 & 2.18  & \xmark & BMC3 & \textbf{1.85} & \xmark & BMC3 & 1.97 & \xmark\\
F & Napnet  &6& BMC2  & 4.17 & \xmark & BMC2 & 5.22 & \xmark & BMC3 & \textbf{1.48} & \xmark\\
&\(\cdots\)& & \multicolumn{3}{c|}{\(\cdots\)} & \multicolumn{3}{c|}{\(\cdots\)} & \multicolumn{3}{c}{\(\cdots\)}\\
F & Fatman &17& BMC3 & 168.78 & \xmark & BMC3 & 169.82 & \xmark  & BMC3 & \textbf{6.72} & \xmark\\
&\(\cdots\)& & \multicolumn{3}{c|}{\(\cdots\)} & \multicolumn{3}{c|}{\(\cdots\)} & \multicolumn{3}{c}{\(\cdots\)}\\
F & Belnet2009 &21& BMC2  & 1146.26 & \xmark & BMC2 & 611.81 & \xmark  & BMC3 & \textbf{24.26} & \xmark\\
&\(\cdots\)& & \multicolumn{3}{c|}{\(\cdots\)} & \multicolumn{3}{c|}{\(\cdots\)} & \multicolumn{3}{c}{\(\cdots\)}\\
F & KentmanJan2011 &38& BMC3 & 167.92 & \xmark & BMC3 & 86.44 & \xmark  & BMC2 & \textbf{9.35} & \xmark\\
&\(\cdots\)& & \multicolumn{3}{c|}{\(\cdots\)} & \multicolumn{3}{c|}{\(\cdots\)} & \multicolumn{3}{c}{\(\cdots\)}\\
F & Latnet &69& - & TO & ? & - & TO & ? & BMC2 & \textbf{209.20} & \xmark\\
F & Ulaknet &82& - & TO & ? & - & TO & ? & BMC2 & \textbf{1043.74} & \xmark\\\hline\hline
\multicolumn{3}{l|}{Sum of runtimes (in hours):}&\multicolumn{3}{r|}{82.99}&\multicolumn{3}{r|}{79.15}&\multicolumn{3}{r}{30.31}\\
\multicolumn{3}{l|}{Nb of TOs (of 230 exper.):}&\multicolumn{3}{r|}{146}&\multicolumn{3}{r|}{138}&\multicolumn{3}{r}{6}
 \end{tabular}
\vspace{-0.4cm}
\end{table}
}

We conduct a case study based on SDN with a corresponding artifact~\cite{GiesekingH19}. 
The performance improvements of \adamMC{} compared to the prototype~\cite{DBLP:conf/atva/FinkbeinerGHO19} are summarized in Table~\ref{table:benchmarks}.  
For realistic software-defined networks~\cite{DBLP:journals/jsac/KnightNFBR11}, one ingress and one egress switch are chosen at random. 
Two forwarding tables between the two switches and an update from the first to the second configuration are chosen at random. 
\adamMC{} verifies that the update maintained \emph{connectivity} 
between ingress and egress switch. 
The results are depicted in rows starting with~T. 
For rows starting with~F, we required \emph{connectivity} of a random switch which is not in the forwarding tables.  
\adamMC{} falsified this requirement for the update. 

The prototype implementation based on an \emph{explicit encoding} can verify updates of networks with 7 switches and falsify updates of networks with 38 switches.
We optimize the explicit encoding to a \emph{logarithmic encoding} and the number of switches for which updates can be verified increases to~17.
More significantly, the \emph{parallel approach} in combination with the logarithmic encoding leads to tremendous performance gains.
The performance gains of an approach with inferior worst-case complexity are mainly due to the
smaller complexity of the LTL formula created by the reduction. 
The encoding of SDN requires fairness assumptions for each transition.
These assumptions (encoded in the run part of the formula) experience a blow-up with until operators
by the sequential approach but only need a disjunction in the parallel approach.
Hence, the size of networks for which \adamMC{} can verify updates increases to 38~switches
and the size for which it can falsify updates increases to 82 switches. 
For rather small networks, the tool needs only a few seconds to verify and falsify updates
which makes it a great option for operators when updating networks.

\section{Related Work}

We refer to~\cite{DBLP:journals/pieee/KreutzRVRAU15} for an introduction to SDN.
Solutions for correctness of updates of software-defined networks include \emph{consistent
updates}~\cite{DBLP:conf/sigcomm/ReitblattFRSW12,DBLP:conf/wdag/CernyFJM16}, \emph{dynamic scheduling}~\cite{DBLP:conf/sigcomm/JinLGKMZRW14}, and \emph{incremental updates}~\cite{DBLP:conf/sigcomm/KattaRW13}.
Both explicit and SMT-based model checking~\cite{DBLP:conf/nsdi/CaniniVPKR12,DBLP:conf/fmcad/MajumdarTW14,DBLP:conf/sigcomm/MaiKACGK11,DBLP:conf/icnp/WangMLTS13,DBLP:conf/pldi/BallBGIKSSV14,DBLP:conf/popl/PadonIKLSS15} is used to verify software-defined
networks.
Closest
to our approach are models of networks as Kripke structures 
to use model checking for synthesis of correct network
updates~\cite{DBLP:conf/cav/El-HassanyTVV17,DBLP:conf/cav/McClurgHC17}. 
The model checking
subroutine of the synthesizer assumes that each packet sees at most
one updated switch.
Our model checking routine does not make such an assumption.

There is a significant number of model checking tools~(e.g., \cite{DBLP:conf/apn/Wolf18a,DBLP:conf/tacas/Thierry-Mieg15}) 
for Petri nets and an annual model checking contest~\cite{mcc:2019}. 
\adamMC{} is restricted to safe Petri nets whereas other tools can handle bounded and colored Petri nets.
At the same time, only \adamMC{} accepts LTL formulas with places \emph{and} transitions as atomic propositions.
This is essential to express fairness in our SDN encoding.

\section{Conclusion}

We presented the tool \adamMC{} with its three application domains: checking safe Petri nets with transits against Flow-LTL, checking concurrent updates of software-defined networks against common assumptions and specifications, and checking safe Petri nets against LTL. 
New algorithms allow \adamMC{} to model check software-defined networks of realistic size: it can verify updates of networks with up to 38 switches and can falsify updates of networks with up to 82 switches. 

\bibliographystyle{splncs04}
\bibliography{ms}

\appendix
\section*{Appendix}
\section{Technical Details}
\label{app:parApp}
In this part of the appendix, details of the parallel approach, i.e.,
the construction of the Petri net \(\pNetMC\) and the construction of the LTL formula~\(\phiMC\),
are given.
\subsection{Construction of the Net Transformation (Parallel Approach)}
Let \(\netIds\) be a set of unique identifiers and
\(\netId_\pNet:\pl\cup\tr\to\netIds\) an injective naming function
which uniquely identifies every place and transition of a given Petri net~\(\pNet\) (or of a Petri net with transits).
We omit the subscript if the net is clear from the context.
To keep the presentation clear, we often directly use \(\mathit{identifier}\)
for a node \(n\in\pl\cup\tr\) with \(\netId(n)=\mathit{identifier}\).

The construction of a Petri net with transits to a standard P/T Petri net with inhibitor arcs
is given by the following definition.
\begin{definition}[Petri Net with Transits to a P/T Petri Net]
\label{def:pnwt2pn}
For a Petri net with transits \(\petriNetFl\)
and a Flow-LTL formula~\(\phiRun\)
with \(n\) subformulas, a Petri net \(\pNetMC=(\plMC,\trMC,\flMC,\inhibitorFlMC,\initMC)\)
with inhibitor arcs (denoted by \(\inhibitorFlMC\))
and a labeling function \(\lambda: \trMC\to\tr\)
are defined as follows:
\begin{itemize}
\item[(p)] The places of the original net \(\pNet\) are copied \(n+1\) times:
\[\plMC=\pl\cup\bigcup_{\{1,\ldots,n\}} \left(\{\initsub_i\} \cup \{\subnet{p}_i\with p\in\pl\}\right)\]
\item[(t)] For every transition \(t\in\tr\) and every combination of which subnet is tracking which transit (or no transit with marker \(\noTransits\)), 
there is one transition in \(\pNetMC\).
Each transition is connected to the original part of the net according to \(t\).
For the subnet part, it either a) moves the token from the initial place according to the transit,
b) moves the token from the corresponding ingoing place of the transit according to the transit,
or, in the case that the subnet is not involved in any of the transits,
c) is connected by inhibitor arcs to all ingoing places of the transition \(t\).
\begin{align*}
&\forall t\in\tr:\forall \mathfrak{c}=((x_1,p_1),\ldots,(x_n,p_n))\in(\tokenflow(t)\cup\{\noTransits\})^n:\exists t\MC\in\trMC:\\
&\quad\forall (p,t),(t,q)\in\fl: (p,t\MC),(t\MC,q)\in\flMC\wedge\netId(t\MC)=\netId(t)_\mathfrak{c}\wedge\lambda(t\MC)=\netId(t)\,\wedge \\
&\quad\forall i\in \{1,\ldots,n\}: x_i = \startfl\quad\;\; \implies (\initsub_i,t\MC),(t\MC,\subnet{p_i}_i)\in\flMC\, \wedge \\
&\quad\qquad\qquad\qquad\quad\,    x_i,p_i\in\pl \;\implies (\subnet{x_i}_i,t\MC),(t\MC,\subnet{p_i}_i)\in\flMC\, \wedge\\
&\quad\qquad\qquad\qquad\quad    (x_i,p_i)=\noTransits \implies \forall (p,t)\in\fl: (\subnet{p_i}_i,t\MC)\in\inhibitorFlMC
\end{align*}
\item[(I)] The initial marking is given by \(\initMC=\init\cup\{\initsub_i \with i\in\{1,\ldots,n\}\}\).
\end{itemize}
The sets \(\trMC, \flMC\), and \(\inhibitorFlMC\) are defined as the smallest sets fulfilling 
condition~\textit{(t)}.
The identifiers with the square brackets and those with a combination \(\mathfrak{c}\) in their index are fresh identifiers.
\end{definition}
The results regarding the size of the constructed net directly follow from the definition
and that there are \(|\pl|\cdot|\tr|\cdot|\pl|+|\tr|\cdot|\pl|\) transits in the worst-case.

\begin{lemma}[Size of the Constructed Net]
\label{lem:sizePN}
The constructed Petri net \(\pNetMC\) has
\(\oclass{|\pNet|\cdot n + |\pNet|}\) places 
and \(\oclass{|\pNet|^{3n}+|\pNet|}\) transitions.
\end{lemma}

\subsection{Construction of the Formula Transformation (Parallel Approach)}
\begin{figure}[t]
\centering
\begin{tikzpicture}

\draw[thick, -Triangle] (1,0) -- (11cm,0) node[font=\scriptsize,below left=3pt and -8pt]{time steps};

\foreach \x in {1,2,...,10}
\draw[-] (\x cm,3pt) -- (\x cm,-3pt);

\foreach \x in {1,2,...,9}
\draw[lightgray, line width=4pt,-] 
(\x,.5) -- +(1,0);
\draw[-, dashed, lightgray] (10.25,.5) --  +(0.5,0);
\node at (0.6,.5) {\(\tau\)};

\foreach \x in {2,3, 7,9}
\draw[gray, line width=4pt,-] 
(\x,1) -- +(1,0);
\draw[-, dashed, gray] (10.25,1) --  +(0.5,0);
\node at (0.6,1) {\(\beta_1\)};

\node at (0.6,1.85) {\(\vdots\)};

\foreach \x in {1,2,7,8}
\draw[black, line width=4pt,-] 
(\x,2.5) -- +(1,0);
\draw[-, dashed, black] (10.25,2.5) --  +(0.5,0);
\node at (0.6,2.5) {\(\beta_i\)};

\node at (0.6,3.2) {\(\vdots\)};

\end{tikzpicture}
\caption{A possible sequence of the global timeline \(\tau\) and the timelines of the possible infinite number of flow chains
\(\beta_i\). A filled time step for a timeline of a flow chain indicates that the fired transition has a transit which extends this flow chain.
\vspace{-0.5cm}
}
	\label{fig:timelines}
\end{figure}
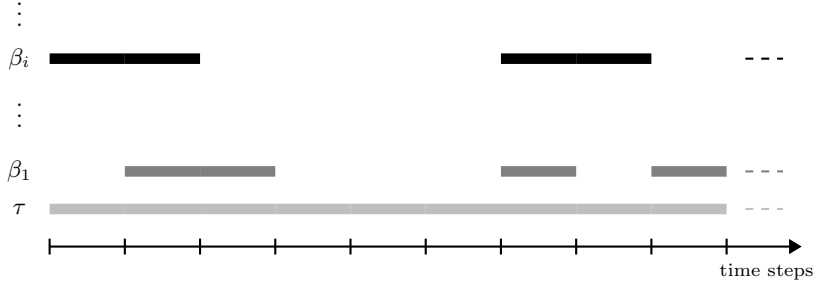

We create an LTL formula \(\phiMC\) to the Petri net \(\pNetMC\)
(created by Def.~\ref{def:pnwt2pn} of a Petri net with transits \(\petriNetFl\))
from a Flow-LTL formula~\(\phiRun\) with \(n\in\N\) flow subformulas \(\varphi_{F_i}=\A\,\phiLTL_i\).
The intricate part of the construction is to deal with the different timelines.
On the one hand, there is the global timeline of the Petri net \(\pNet\).
This timeline can be used to check the run part of the formula.
On the other hand, there are the different timelines of the possible infinite flow chains.
For the flow chains, the global steps not concerning the chain have to be adequately skipped with until operators.
Figure~\ref{fig:timelines} gives an overview of a possible sequence of different timelines.

We define the set of transitions tracking a chain of a specific subnet \(i\in\{1,\ldots,n\}\)
by \(\trMC_i=\{t\in\trMC\with \exists p\in\pl: (\subnet{p}_i,t)\in\flMC \vee (t,\subnet{p}_i)\in\flMC\}\)
and the set of all other transitions by \(O_i=\trMC\setminus\trMC_i\).
For a transition \(t\in\tr\), the set \(M_i(t)=\trMC_i\cap\{t\MC\in\trMC\with\lambda(t\MC)=t\}\)
collects all corresponding transitions tracking a chain of the subnet.

First, the \emph{places} of the \emph{flow subformulas} have to be substituted by the corresponding places tracking the chain,
i.e., all occurrences of a place \(p\in\pl\) in a flow subformula \(\varphi_{F_i}\) are simultaneously replaced by \(\subnet{p}_i\).
Second, the \emph{transitions} of the flow subformulas have to be substituted such that all steps of the global timeline
which do not involve the tracked flow chain are skipped until a transition involving the flow chain is fired, i.e.,
all occurrences of a transition \(t\in\tr\) in a flow subformula \(\varphi_{F_i}\) are simultaneously substituted
by \((\bigvee_{t_o\in O_i} t_o)\U (\bigvee_{t_m\in M_i(t)}t_m)\).
Similarly, the \emph{next operator} of the flow subformulas have to be substituted such that the steps of the global timeline
are skipped until a step involving the tracking subnet is taken.
Here two cases have to be considered: either the chain ends, i.e., no transition of the subnet is ever fired again,
then the formula has to directly hold in the stuttering part, or there is a transition of the subnet, then the formula has to hold
in the direct successor state. 
This means all occurrences of a subformula \(\ltlNext\,\phi\) in a flow subformula \(\varphi_{F_i}\) are replaced from the inner-
to the outermost occurrence by \(((\bigvee_{t\in O_i} t)\U ((\bigvee_{t\in\trMC_i}t)\wedge\ltlNext\,\phi))\vee (\ltlAlways\, (\neg(\bigvee_{t\in \trMC_i}t))\wedge\phi)\).

For the \emph{run part} of the formula, we can directly use the global timeline, i.e., the \emph{next operator} needs no substitution.
Further, the \emph{places} are already correctly named.
Only the \emph{transitions} \(t\in\tr\) in the run part of \(\phiRun\) have to be substituted simultaneously
by \(\bigvee_{t'\in{}\{t'\in\trMC\with \lambda(t')=t\}} t'\) to allow for all transitions corresponding to \(t\).

Finally, the flow subformulas are simultaneously substituted by \(\initsub_i\LTLweakuntil(\neg \initsub_i\wedge \phiLTL_i')\)
(where \(\phiLTL_i'\) is the result of the above mentioned substitutions within a flow subformula)
such that all steps of the global timeline are skipped until a flow chain is created and tracked.
Table~\ref{tab:formulas} gives an overview of these substitutions.

\vspace{-0.75cm}
\def\arraystretch{1.1}%
\setlength\tabcolsep{1.2mm}
\begin{table}
\caption{An overview of the necessary substitutions to create \(\phiMC\) from \(\phiRun\).
The next operator is substituted from the innermost to the outermost occurrence, the other subformulas are substituted simultaneously.
} 
\label{tab:formulas}
\centering
\begin{tabular}{l|l|l}
 & Run part of \(\phiRun\) & Flow subformula \(\A\,\phiLTL_i\) part of \(\phiRun\) \\\hline
\(p\in\pl\) & \(p\) & \(\subnet{p}_i\) \\
\(t\in\tr\)  & \(\bigvee_{t'\in{}\{t'\in\trMC\with \lambda(t')=t\}} t'\) & \((\bigvee_{t_o\in O_i} t_o)\U (\bigvee_{t_m\in M_i(t)}t_m)\) \\
\(\ltlNext\,\phi\) & \(\ltlNext\,\phi\) &  \(((\bigvee_{t\in O_i} t)\U ((\bigvee_{t\in\trMC_i}t)\wedge\ltlNext\,\phi))\vee (\ltlAlways\, (\neg(\bigvee_{t\in \trMC_i}t))\wedge\phi)\)\\
\(\A\,\phiLTL_i\)  & \(\initsub_i\LTLweakuntil(\neg \initsub_i\wedge \phiLTL_i')\) & --
\end{tabular}
\end{table}
\vspace{-0.5cm}

The size of the constructed formula directly results from the blow-up of the number of transition during the creation of \(\pNetMC\)
and the substitutions introducing the disjunctions over these transition in the creation of \(\phiMC\).

\begin{lemma}[Size of the Constructed Formula]
\label{lem:sizeFormula}
The size of the constructed LTL formula \(\phiMC\) is in \(\oclass{|\pNet|^{3n} \cdot |\phiRun| + |\phiRun|}\).
\end{lemma}

Note that there is only a significant blow-up in the formula when transitions are used as atomic propositions 
in either the flow or the run part of the formula or when the next operator is used in the flow part of the formula.
Moreover, even the usage of transitions as atomic propositions in the run part of the formula only results in
a blow-up by all combinations of transits of this transition regarding the subnets. 
In practical applications, this makes a huge difference compared to the sequential approach,
because
model checking is exponential in the size of the formula
and many examples need fairness assumptions, i.e., transitions in the run part of the formula,
and have only few local requirements.

The proof of the correctness of the transformations for the parallel approach is very similar
to the one of the sequential approach presented in \cite{DBLP:journals/corr/abs-1907-11061}.
We again can mutually transform the counterexample to show the contraposition
\(\pNet\not\models\phiRun\text{ iff }\pNetMC\not\models_\mathtt{LTL}\phiMC\).
Here we do not have to pump up the firing sequence serving as counterexample for \(\pNet\models\phiRun\),
but have to replace each transition by a transition which adequately extends all flow chains of the counterexample.
For the other direction, we can replace the transitions of the counterexample by the labels of the transitions 
and, analog to the sequential approach, iteratively concatenate the transitions and places of the subnets to 
gain the flow chains serving as counterexamples for the subformula part.
The complicated parts of the structural induction, i.e., adequately skipping the global time steps
for the flow subformulas, can be done analogously because the formulas of the parallel approach and the sequential approach
are similar in this case and fit to the different structure of the net.

\section{Complete Results}
\setlength{\tabcolsep}{1pt}
\begin{longtable}{ccc!{\vline}crc!{\vline}crc!{\vline}crc}
\caption{We compare the explicit and the logarithmic encoding of the sequential approach with the parallel approach. The results are the average over 5 runs from an Intel i7-2700K CPU with 3.50~GHz, 32~GB RAM, and a timeout of 30~minutes. We report the runtimes of IC3 to verify (T) updates of software-defined networks and the runtimes of both BMC2 and BMC3 to falsify (F) updates of software-defined networks all with respect to connectivity between randomly chosen ingress and egress switches and forwarding tables.}\\
  \label{table:FullBenchmarks}
  \centering
 & & & \multicolumn{3}{c!{\vline}}{expl.\ enc.} & \multicolumn{3}{c!{\vline}}{log.\ enc.} & \multicolumn{3}{c}{parallel appr.}\\
T / F & Network & \(\#\mathit{Sw}\) & Alg. & Time & \(\models\) & Alg. & Time & \(\models\) & Alg. & Time & \(\models\)\\
 \hline
\rowcolor[gray]{.90}T&Arpanet196912&4&IC3&12.0760&\cmark&IC3&9.8872&\cmark&IC3&2.1760&\cmark\\
\rowcolor[gray]{.90}T&Napnet&6&IC3&146.4920&\cmark&IC3&96.0640&\cmark&IC3&4.7448&\cmark\\
T&Epoch&6&IC3&240.5720&\cmark&IC3&214.6960&\cmark&IC3&6.7800&\cmark\\
T&Telecomserbia&6&IC3&1182.4320&\cmark&IC3&912.7560&\cmark&IC3&12.1232&\cmark\\
T&Layer42&6&IC3&133.1992&\cmark&IC3&131.6824&\cmark&IC3&6.2624&\cmark\\
T&Dataxchange&6&-&TO&?&IC3&380.1976&\cmark&IC3&19.9968&\cmark\\
T&Sanren&7&IC3&304.6368&\cmark&IC3&437.0128&\cmark&IC3&16.6776&\cmark\\
T&Getnet&7&IC3&940.4160&\cmark&IC3&103.0960&\cmark&IC3&11.0480&\cmark\\
T&Netrail&7&IC3&171.5952&\cmark&IC3&531.5576&\cmark&IC3&31.9800&\cmark\\
\rowcolor[gray]{.90}T&Heanet&7&IC3&806.8144&\cmark&IC3&84.6160&\cmark&IC3&30.3008&\cmark\\
\rowcolor[gray]{.90}T&HiberniaIreland&7&-&TO&?&-&TO&?&IC3&26.5824&\cmark\\
\rowcolor[gray]{.90}T&Arpanet19706&9&-&TO&?&IC3&362.2056&\cmark&IC3&11.3304&\cmark\\
\rowcolor[gray]{.90}T&Nordu2005&9&-&TO&?&-&TO&?&IC3&12.6688&\cmark\\
T&Nsfcnet&10&-&TO&?&-&TO&?&IC3&5.5448&\cmark\\
T&Sprint&11&-&TO&?&-&TO&?&IC3&745.7408&\cmark\\
T&TLex&12&-&TO&?&-&TO&?&IC3&17.1296&\cmark\\
T&Compuserve&13&-&TO&?&-&TO&?&IC3&107.9464&\cmark\\
T&Eenet&13&-&TO&?&-&TO&?&IC3&40.3456&\cmark\\
T&HiberniaCanada&13&-&TO&?&-&TO&?&IC3&107.6000&\cmark\\
T&Navigata&13&-&TO&?&-&TO&?&IC3&360.5248&\cmark\\
T&Nsfnet&13&-&TO&?&-&TO&?&IC3&181.3240&\cmark\\
T&Uninet&13&-&TO&?&-&TO&?&IC3&1336.1420&\cmark\\
T&Eunetworks&14&-&TO&?&-&TO&?&IC3&80.3952&\cmark\\
T&Ilan&14&-&TO&?&-&TO&?&IC3&137.8408&\cmark\\
T&Claranet&15&-&TO&?&-&TO&?&IC3&128.0024&\cmark\\
T&HiberniaUk&15&-&TO&?&-&TO&?&IC3&184.5888&\cmark\\
T&Spiralight&15&-&TO&?&-&TO&?&IC3&153.2312&\cmark\\
T&Garr199901&16&-&TO&?&-&TO&?&IC3&164.5248&\cmark\\
T&KentmanJul2005&16&-&TO&?&-&TO&?&IC3&163.2448&\cmark\\
T&Marwan&16&-&TO&?&-&TO&?&IC3&136.5992&\cmark\\
T&Peer1&16&-&TO&?&-&TO&?&IC3&357.0224&\cmark\\
T&Rhnet&16&-&TO&?&-&TO&?&IC3&62.6520&\cmark\\
\rowcolor[gray]{.90}T&Fatman&17&-&TO&?&IC3&1543.3360&\cmark&IC3&162.1672&\cmark\\
T&Nextgen&17&-&TO&?&-&TO&?&IC3&403.3296&\cmark\\
T&Nordu2010&18&-&TO&?&-&TO&?&IC3&50.1136&\cmark\\
T&Pacificwave&18&-&TO&?&-&TO&?&IC3&932.5960&\cmark\\
T&Ans&18&-&TO&?&-&TO&?&IC3&1511.3020&\cmark\\
T&Arpanet19719&18&-&TO&?&-&TO&?&IC3&840.6400&\cmark\\
T&BsonetEurope&18&-&TO&?&-&TO&?&IC3&496.2936&\cmark\\
T&HiberniaNireland&18&-&TO&?&-&TO&?&IC3&229.0768&\cmark\\
T&Noel&19&-&TO&?&-&TO&?&IC3&402.8256&\cmark\\
T&Restena&19&-&TO&?&-&TO&?&IC3&698.4024&\cmark\\
T&Savvis&19&-&TO&?&-&TO&?&IC3&1382.3480&\cmark\\
T&Twaren&20&-&TO&?&IC3&1167.9080&\cmark&IC3&1388.6660&\cmark\\
T&Janetlense&20&-&TO&?&-&TO&?&IC3&730.6448&\cmark\\
T&BtAsiaPac&20&-&TO&?&-&TO&?&IC3&1311.1100&\cmark\\
T&Oxford&20&-&TO&?&-&TO&?&IC3&678.3344&\cmark\\
T&Harnet&21&-&TO&?&-&TO&?&IC3&347.0536&\cmark\\
T&Belnet2009&21&-&TO&?&-&TO&?&IC3&1604.5860&\cmark\\
T&GtsRomania&21&-&TO&?&-&TO&?&IC3&236.7912&\cmark\\
T&Packetexchange&21&-&TO&?&-&TO&?&IC3&688.5176&\cmark\\
T&Garr200404&22&-&TO&?&-&TO&?&IC3&184.8440&\cmark\\
T&Belnet2010&22&-&TO&?&-&TO&?&IC3&346.6064&\cmark\\
T&Garr200109&22&-&TO&?&-&TO&?&IC3&1499.7620&\cmark\\
T&KentmanApr2007&22&-&TO&?&-&TO&?&IC3&429.7848&\cmark\\
T&Istar&23&-&TO&?&-&TO&?&IC3&169.0848&\cmark\\
T&Garr199905&23&-&TO&?&-&TO&?&IC3&440.1864&\cmark\\
T&Garr199904&23&-&TO&?&-&TO&?&IC3&590.0240&\cmark\\
T&Cesnet2001&23&-&TO&?&-&TO&?&IC3&308.9808&\cmark\\
T&Fccn&23&-&TO&?&-&TO&?&IC3&752.9816&\cmark\\
T&Uran&24&-&TO&?&-&TO&?&IC3&82.6056&\cmark\\
T&Garr200112&24&-&TO&?&-&TO&?&IC3&1731.2440&\cmark\\
T&Psinet&24&-&TO&?&-&TO&?&IC3&226.0680&\cmark\\
T&Arpanet19723&25&-&TO&?&-&TO&?&IC3&218.3872&\cmark\\
T&Vinaren&25&-&TO&?&-&TO&?&IC3&1084.5340&\cmark\\
T&KentmanFeb2008&26&-&TO&?&-&TO&?&IC3&421.7424&\cmark\\
T&Garr200212&27&-&TO&?&-&TO&?&IC3&1205.4020&\cmark\\
T&Bbnplanet&27&-&TO&?&-&TO&?&IC3&896.9776&\cmark\\
T&Darkstrand&27&-&TO&?&-&TO&?&IC3&1466.4960&\cmark\\
T&KentmanAug2005&28&-&TO&?&-&TO&?&IC3&278.9248&\cmark\\
\rowcolor[gray]{.90}T&Myren&37&-&TO&?&-&TO&?&IC3&1309.2280&\cmark\\
\rowcolor[gray]{.90}T&KentmanJan2011&38&-&TO&?&-&TO&?&IC3&1261.3220&\cmark\\
\cline{1-12}
F&Arpanet196912&4&BMC2&2.3528&\xmark&BMC2&2.0952&\xmark&BMC2&1.1992&\xmark\\
\rowcolor[gray]{.90}F&Arpanet196912&4&BMC3&2.1768&\xmark&BMC3&1.8528&\xmark&BMC3&1.1968&\xmark\\
\rowcolor[gray]{.90}F&Napnet&6&BMC2&4.1688&\xmark&BMC2&5.2240&\xmark&BMC2&1.6408&\xmark\\
\rowcolor[gray]{.90}F&Napnet&6&BMC3&5.7072&\xmark&BMC3&5.4368&\xmark&BMC3&1.4808&\xmark\\
F&Epoch&6&BMC2&20.7584&\xmark&BMC2&14.3200&\xmark&BMC2&2.6328&\xmark\\
F&Epoch&6&BMC3&15.4112&\xmark&BMC3&13.5632&\xmark&BMC3&2.3912&\xmark\\
F&Telecomserbia&6&BMC2&45.1120&\xmark&BMC2&39.7160&\xmark&BMC2&13.2600&\xmark\\
F&Telecomserbia&6&BMC3&37.5104&\xmark&BMC3&41.4688&\xmark&BMC3&12.8704&\xmark\\
F&Layer42&6&BMC2&9.4880&\xmark&BMC2&11.8768&\xmark&BMC2&2.0744&\xmark\\
F&Layer42&6&BMC3&11.4400&\xmark&BMC3&6.7544&\xmark&BMC3&2.5560&\xmark\\
F&Sanren&7&BMC2&64.8976&\xmark&BMC2&134.8184&\xmark&BMC2&8.2312&\xmark\\
F&Sanren&7&BMC3&173.9256&\xmark&BMC3&81.2960&\xmark&BMC3&3.8832&\xmark\\
F&Getnet&7&BMC2&7.3792&\xmark&BMC2&9.2480&\xmark&BMC2&1.6872&\xmark\\
F&Getnet&7&BMC3&7.5144&\xmark&BMC3&7.2872&\xmark&BMC3&1.5248&\xmark\\
F&Netrail&7&BMC2&80.8968&\xmark&BMC2&50.0872&\xmark&BMC2&5.6976&\xmark\\
F&Netrail&7&BMC3&63.8416&\xmark&BMC3&72.6552&\xmark&BMC3&3.5632&\xmark\\
F&Heanet&7&BMC2&57.2528&\xmark&BMC2&66.6016&\xmark&BMC2&5.6632&\xmark\\
F&Heanet&7&BMC3&54.5128&\xmark&BMC3&27.9272&\xmark&BMC3&5.5848&\xmark\\
F&Arpanet19706&9&BMC2&52.5888&\xmark&BMC2&44.3200&\xmark&BMC2&7.7576&\xmark\\
F&Arpanet19706&9&BMC3&58.1392&\xmark&BMC3&33.6264&\xmark&BMC3&4.5640&\xmark\\
F&Nordu2005&9&BMC2&35.9496&\xmark&BMC2&33.2320&\xmark&BMC2&6.0872&\xmark\\
F&Nordu2005&9&BMC3&38.6664&\xmark&BMC3&25.6184&\xmark&BMC3&3.2904&\xmark\\
F&Nsfcnet&10&BMC2&14.3520&\xmark&BMC2&13.2688&\xmark&BMC2&2.1520&\xmark\\
F&Nsfcnet&10&BMC3&13.5312&\xmark&BMC3&4.8568&\xmark&BMC3&2.0184&\xmark\\
F&Sprint&11&-&TO&?&-&TO&?&BMC2&567.1048&\xmark\\
F&Sprint&11&-&TO&?&-&TO&?&BMC3&582.5752&\xmark\\
F&TLex&12&BMC2&92.9472&\xmark&BMC2&81.4936&\xmark&BMC2&4.9240&\xmark\\
F&TLex&12&BMC3&89.6536&\xmark&BMC3&49.1896&\xmark&BMC3&7.0464&\xmark\\
F&Compuserve&13&-&TO&?&-&TO&?&BMC2&544.4312&\xmark\\
F&Compuserve&13&-&TO&?&-&TO&?&BMC3&460.1632&\xmark\\
F&Eenet&13&BMC2&249.1056&\xmark&BMC2&238.0224&\xmark&BMC2&21.8432&\xmark\\
F&Eenet&13&BMC3&271.4032&\xmark&BMC3&218.4312&\xmark&BMC3&19.5312&\xmark\\
F&HiberniaCanada&13&BMC2&1309.4440&\xmark&BMC2&935.0480&\xmark&BMC2&149.3392&\xmark\\
F&HiberniaCanada&13&BMC3&1304.8740&\xmark&BMC3&1097.0560&\xmark&BMC3&53.5256&\xmark\\
F&Navigata&13&-&TO&?&-&TO&?&BMC2&440.9632&\xmark\\
F&Navigata&13&-&TO&?&-&TO&?&BMC3&319.2552&\xmark\\
F&Nsfnet&13&-&TO&?&-&TO&?&BMC2&159.0920&\xmark\\
F&Nsfnet&13&-&TO&?&BMC3&1621.8300&\xmark&BMC3&177.6008&\xmark\\
F&Eunetworks&14&BMC2&1039.4640&\xmark&BMC2&1104.7060&\xmark&BMC2&38.2784&\xmark\\
F&Eunetworks&14&BMC3&1417.8800&\xmark&BMC3&1056.0080&\xmark&BMC3&35.9568&\xmark\\
F&Claranet&15&BMC2&189.4912&\xmark&BMC2&165.1504&\xmark&BMC2&10.2744&\xmark\\
F&Claranet&15&BMC3&160.3808&\xmark&BMC3&150.6000&\xmark&BMC3&7.9720&\xmark\\
F&Spiralight&15&-&TO&?&-&TO&?&BMC2&1249.4900&\xmark\\
F&Spiralight&15&-&TO&?&-&TO&?&BMC3&1734.8840&\xmark\\
F&Garr199901&16&BMC2&625.8648&\xmark&BMC2&432.1856&\xmark&BMC2&31.3360&\xmark\\
F&Garr199901&16&BMC3&743.3096&\xmark&BMC3&370.4792&\xmark&BMC3&61.6488&\xmark\\
F&KentmanJul2005&16&BMC2&1391.2280&\xmark&BMC2&1243.6260&\xmark&BMC2&175.4600&\xmark\\
F&KentmanJul2005&16&BMC3&1405.8180&\xmark&BMC3&1030.0400&\xmark&BMC3&171.0152&\xmark\\
F&Marwan&16&-&TO&?&-&TO&?&BMC2&696.8736&\xmark\\
F&Marwan&16&-&TO&?&-&TO&?&BMC3&799.7816&\xmark\\
F&Peer1&16&-&TO&?&-&TO&?&-&TO&?\\
F&Peer1&16&-&TO&?&-&TO&?&BMC3&1551.8120&\xmark\\
F&Rhnet&16&-&TO&?&-&TO&?&BMC2&105.4688&\xmark\\
F&Rhnet&16&-&TO&?&-&TO&?&BMC3&49.9360&\xmark\\
F&Fatman&17&BMC2&193.7456&\xmark&BMC2&200.3976&\xmark&BMC2&18.1704&\xmark\\
\rowcolor[gray]{.90}F&Fatman&17&BMC3&168.7768&\xmark&BMC3&169.8224&\xmark&BMC3&6.7232&\xmark\\
F&Goodnet&17&-&TO&?&-&TO&?&BMC2&410.3936&\xmark\\
F&Goodnet&17&-&TO&?&-&TO&?&BMC3&378.1480&\xmark\\
F&Nextgen&17&-&TO&?&-&TO&?&-&TO&?\\
F&Nextgen&17&-&TO&?&-&TO&?&BMC3&5014.4240&\xmark\\
F&Nordu2010&18&BMC2&183.4608&\xmark&BMC2&140.1384&\xmark&BMC2&13.6824&\xmark\\
F&Nordu2010&18&BMC3&116.1192&\xmark&BMC3&80.7856&\xmark&BMC3&6.9120&\xmark\\
F&Pacificwave&18&BMC2&1761.4720&\xmark&BMC2&1035.8420&\xmark&BMC2&95.0104&\xmark\\
F&Pacificwave&18&BMC3&1166.0460&\xmark&BMC3&1545.4200&\xmark&BMC3&64.5768&\xmark\\
F&BsonetEurope&18&-&TO&?&-&TO&?&BMC2&770.6568&\xmark\\
F&BsonetEurope&18&-&TO&?&-&TO&?&BMC3&1195.4540&\xmark\\
F&Highwinds&18&-&TO&?&-&TO&?&BMC2&671.2392&\xmark\\
F&Highwinds&18&-&TO&?&-&TO&?&BMC3&1045.0800&\xmark\\
F&Noel&19&-&TO&?&-&TO&?&BMC2&1356.7820&\xmark\\
F&Noel&19&-&TO&?&-&TO&?&BMC3&611.0792&\xmark\\
F&Restena&19&-&TO&?&-&TO&?&BMC2&630.9496&\xmark\\
F&Restena&19&-&TO&?&-&TO&?&BMC3&192.2696&\xmark\\
F&Twaren&20&BMC2&560.8456&\xmark&BMC2&429.7516&\xmark&BMC2&66.7208&\xmark\\
F&Twaren&20&BMC3&650.5672&\xmark&BMC3&349.0816&\xmark&BMC3&31.4976&\xmark\\
F&Marnet&20&BMC2&858.9728&\xmark&BMC2&557.1536&\xmark&BMC2&18.9960&\xmark\\
F&Marnet&20&BMC3&846.0584&\xmark&BMC3&475.7168&\xmark&BMC3&26.4488&\xmark\\
F&Janetlense&20&BMC2&735.1432&\xmark&BMC2&721.8584&\xmark&BMC2&28.1296&\xmark\\
F&Janetlense&20&BMC3&492.5848&\xmark&BMC3&616.7248&\xmark&BMC3&28.6504&\xmark\\
F&BtAsiaPac&20&-&TO&?&-&TO&?&BMC2&1298.5280&\xmark\\
F&BtAsiaPac&20&-&TO&?&-&TO&?&BMC3&897.9200&\xmark\\
F&Oxford&20&-&TO&?&-&TO&?&BMC2&645.1712&\xmark\\
F&Oxford&20&-&TO&?&-&TO&?&BMC3&521.4232&\xmark\\
F&Harnet&21&BMC2&961.5872&\xmark&BMC2&873.2056&\xmark&BMC2&60.4784&\xmark\\
F&Harnet&21&BMC3&1410.2540&\xmark&BMC3&735.7256&\xmark&BMC3&58.4040&\xmark\\
\rowcolor[gray]{.90}F&Belnet2009&21&BMC2&1146.2600&\xmark&BMC2&611.8096&\xmark&BMC2&43.8104&\xmark\\
\rowcolor[gray]{.90}F&Belnet2009&21&-&TO&?&BMC3&1294.5020&\xmark&BMC3&24.2568&\xmark\\
F&Garr200404&22&BMC2&61.5632&\xmark&BMC2&70.9440&\xmark&BMC2&6.1848&\xmark\\
F&Garr200404&22&BMC3&45.9016&\xmark&BMC3&49.3768&\xmark&BMC3&4.6952&\xmark\\
F&Bandcon&22&-&TO&?&-&TO&?&-&TO&?\\
F&Bandcon&22&-&TO&?&-&TO&?&BMC3&1619.6780&\xmark\\
F&KentmanApr2007&22&-&TO&?&-&TO&?&BMC2&914.6352&\xmark\\
F&KentmanApr2007&22&-&TO&?&-&TO&?&BMC3&414.7120&\xmark\\
F&Istar&23&BMC2&574.5056&\xmark&BMC2&302.8448&\xmark&BMC2&29.0296&\xmark\\
F&Istar&23&BMC3&221.1632&\xmark&BMC3&236.4752&\xmark&BMC3&22.8224&\xmark\\
F&Garr199905&23&BMC2&188.2176&\xmark&BMC2&155.1984&\xmark&BMC2&13.4320&\xmark\\
F&Garr199905&23&BMC3&272.3944&\xmark&BMC3&78.0928&\xmark&BMC3&10.9904&\xmark\\
F&Garr199904&23&BMC2&478.9360&\xmark&BMC2&342.0088&\xmark&BMC2&65.7056&\xmark\\
F&Garr199904&23&BMC3&304.7032&\xmark&BMC3&385.6456&\xmark&BMC3&33.4232&\xmark\\
F&Aconet&23&-&TO&?&-&TO&?&BMC2&366.4816&\xmark\\
F&Aconet&23&-&TO&?&-&TO&?&BMC3&642.5488&\xmark\\
F&Belnet2003&23&-&TO&?&-&TO&?&BMC2&135.0384&\xmark\\
F&Belnet2003&23&-&TO&?&-&TO&?&BMC3&126.2848&\xmark\\
F&Belnet2005&23&-&TO&?&-&TO&?&BMC2&795.6688&\xmark\\
F&Belnet2005&23&-&TO&?&-&TO&?&BMC3&558.2600&\xmark\\
F&Cesnet2001&23&-&TO&?&BMC2&1439.4880&\xmark&BMC2&277.3880&\xmark\\
F&Cesnet2001&23&-&TO&?&BMC3&1508.9240&\xmark&BMC3&169.1864&\xmark\\
F&Fccn&23&-&TO&?&-&TO&?&BMC2&772.9776&\xmark\\
F&Fccn&23&-&TO&?&-&TO&?&BMC3&785.8392&\xmark\\
F&Uran&24&BMC2&87.9600&\xmark&BMC2&126.0488&\xmark&BMC2&4.0680&\xmark\\
F&Uran&24&BMC3&71.7760&\xmark&BMC3&72.4816&\xmark&BMC3&3.8056&\xmark\\
F&BtEurope&24&-&TO&?&-&TO&?&BMC2&5035.2740&\xmark\\
F&BtEurope&24&-&TO&?&-&TO&?&BMC3&5030.4000&\xmark\\
F&Garr200112&24&-&TO&?&-&TO&?&BMC2&538.3072&\xmark\\
F&Garr200112&24&-&TO&?&-&TO&?&BMC3&294.9824&\xmark\\
F&Arpanet19723&25&-&TO&?&-&TO&?&BMC2&1774.0800&\xmark\\
F&Arpanet19723&25&-&TO&?&-&TO&?&BMC3&1079.5160&\xmark\\
F&Vinaren&25&BMC2&1537.2920&\xmark&BMC2&799.1280&\xmark&BMC2&138.3432&\xmark\\
F&Vinaren&25&BMC3&1093.3200&\xmark&BMC3&1171.3140&\xmark&BMC3&93.6536&\xmark\\
F&KentmanFeb2008&26&BMC2&152.6880&\xmark&BMC2&230.8040&\xmark&BMC2&7.1976&\xmark\\
F&KentmanFeb2008&26&BMC3&83.3048&\xmark&BMC3&89.7208&\xmark&BMC3&4.4288&\xmark\\
F&Garr200212&27&BMC2&416.9272&\xmark&BMC2&405.2304&\xmark&BMC2&29.7792&\xmark\\
F&Garr200212&27&BMC3&418.8768&\xmark&BMC3&301.6184&\xmark&BMC3&14.4280&\xmark\\
F&Gambia&28&-&TO&?&-&TO&?&BMC2&916.3864&\xmark\\
F&Gambia&28&-&TO&?&-&TO&?&BMC3&586.2224&\xmark\\
F&KentmanAug2005&28&-&TO&?&-&TO&?&BMC2&799.1960&\xmark\\
F&KentmanAug2005&28&-&TO&?&-&TO&?&BMC3&935.1016&\xmark\\
F&Ernet&30&-&TO&?&-&TO&?&BMC2&462.7160&\xmark\\
F&Ernet&30&-&TO&?&-&TO&?&BMC3&361.0344&\xmark\\
F&WideJpn&30&-&TO&?&-&TO&?&-&TO&?\\
F&WideJpn&30&-&TO&?&-&TO&?&BMC3&336.8296&\xmark\\
F&Iinet&31&BMC2&1380.3760&\xmark&BMC2&1217.9420&\xmark&BMC2&76.8728&\xmark\\
F&Iinet&31&BMC3&1468.8020&\xmark&BMC3&701.2528&\xmark&BMC3&97.9888&\xmark\\
F&CrlNetworkServices&33&-&TO&?&-&TO&?&BMC2&1534.4540&\xmark\\
F&CrlNetworkServices&33&-&TO&?&-&TO&?&BMC3&591.3928&\xmark\\
F&GtsSlovakia&35&-&TO&?&-&TO&?&BMC2&938.1456&\xmark\\
F&GtsSlovakia&35&-&TO&?&-&TO&?&BMC3&1014.7040&\xmark\\
F&Bren&37&-&TO&?&-&TO&?&BMC2&346.8384&\xmark\\
F&Bren&37&-&TO&?&-&TO&?&BMC3&799.3200&\xmark\\
F&Myren&37&BMC2&183.0128&\xmark&BMC2&154.1552&\xmark&BMC2&10.1816&\xmark\\
F&Myren&37&BMC3&142.0280&\xmark&BMC3&75.9296&\xmark&BMC3&12.9468&\xmark\\
\rowcolor[gray]{.90}F&KentmanJan2011&38&BMC2&237.3152&\xmark&BMC2&192.5768&\xmark&BMC2&9.3536&\xmark\\
\rowcolor[gray]{.90}F&KentmanJan2011&38&BMC3&167.9208&\xmark&BMC3&86.4384&\xmark&BMC3&10.5464&\xmark\\
F&Cesnet200511&39&-&TO&?&-&TO&?&BMC2&496.0712&\xmark\\
F&Cesnet200511&39&-&TO&?&-&TO&?&BMC3&249.1456&\xmark\\
F&Litnet&43&-&TO&?&-&TO&?&BMC2&234.4000&\xmark\\
F&Litnet&43&-&TO&?&-&TO&?&BMC3&223.7856&\xmark\\
F&Bellsouth&51&-&TO&?&-&TO&?&BMC2&173.6720&\xmark\\
F&Bellsouth&51&-&TO&?&-&TO&?&BMC3&184.7136&\xmark\\
F&BtLatinAmerica&51&-&TO&?&-&TO&?&-&TO&?\\
F&BtLatinAmerica&51&-&TO&?&-&TO&?&BMC3&1346.7440&\xmark\\
F&Garr201103&58&-&TO&?&-&TO&?&BMC2&164.2792&\xmark\\
F&Garr201103&58&-&TO&?&-&TO&?&BMC3&89.1304&\xmark\\
F&Forthnet&62&-&TO&?&-&TO&?&BMC2&1040.0800&\xmark\\
F&Forthnet&62&-&TO&?&-&TO&?&BMC3&752.8008&\xmark\\
\rowcolor[gray]{.90}F&Latnet&69&-&TO&?&-&TO&?&BMC2&209.1984&\xmark\\
F&Latnet&69&-&TO&?&-&TO&?&BMC3&235.7168&\xmark\\
\rowcolor[gray]{.90}F&Ulaknet&82&-&TO&?&-&TO&?&BMC2&1043.7440&\xmark\\
F&Ulaknet&82&-&TO&?&-&TO&?&-&TO&?\\
\cline{1-12}
\end{longtable}

\end{document}